\documentclass[12pt,preprint]{aastex}

\newcommand{\etal}{et al.}

\newcommand{\zzc}{ZZ~Ceti}
\newcommand{\msun}{${\mathrm{M}}_{\odot}$}

\newcommand{\porb}{${\mathrm{P}}_{\mathrm{orb}}$}

\newcommand{\sdsos}{SDSS1610-0102}
\newcommand{\sdsosf}{SDSS\,J161033.64-010223.3}

\shorttitle{Multi-site Campaign on \sdsos}
\shortauthors{Mukadam \etal}

\begin{document}

\title{Multi-site Observations of Pulsation in the Accreting White Dwarf \sdsosf\ (V386~Ser)}

\author{Anjum S. Mukadam\altaffilmark{1,2}, D. M. Townsley\altaffilmark{3}, B. T. G\"{a}nsicke\altaffilmark{4},
P. Szkody\altaffilmark{1,2}, T. R. Marsh\altaffilmark{4}, E. L. Robinson\altaffilmark{5}, L. Bildsten\altaffilmark{6,7},
A. Aungwerojwit\altaffilmark{4,8},
M. R. Schreiber\altaffilmark{9}, J. Southworth\altaffilmark{4}, A. Schwope\altaffilmark{10}, B.-Q. For\altaffilmark{5},
G. Tovmassian\altaffilmark{11}, S. V. Zharikov\altaffilmark{11}, M. G. Hidas\altaffilmark{12,7,13}, N. Baliber\altaffilmark{12,7},
T. Brown\altaffilmark{12,7}, P. A. Woudt\altaffilmark{14}, B. Warner\altaffilmark{14}, D. O'Donoghue\altaffilmark{15},
D. A. H. Buckley\altaffilmark{15,16}, R. Sefako\altaffilmark{15}, \and E. M. Sion\altaffilmark{17}}

\altaffiltext{1}{Department of Astronomy, University of Washington, Seattle, WA\,-\,98195-1580, USA}
\altaffiltext{2}{Apache Point Observatory, 2001 Apache Point Road, Sunspot, NM 88349-0059}
\altaffiltext{3}{Department of Astronomy, University of Arizona, 933 N. Cherry Av, Tucson, AZ 85721}
\altaffiltext{4}{Department of Physics, University of Warwick, Coventry, CV4 7AL, UK}
\altaffiltext{5}{Department of Astronomy, University of Texas at Austin, Austin, TX - 78712}
\altaffiltext{6}{Kavli Institute for Theoretical Physics, University of California, Santa Barbara, CA 93106}
\altaffiltext{7}{Department of Physics, University of California, Santa Barbara, CA - 93106}
\altaffiltext{8}{Department of Physics, Faculty of Science, Naresuan University, Phitsanulok, 65000, Thailand}
\altaffiltext{9}{Departamento de Fisica y Astronomia, Universidad de Valparaiso, Valparaiso, Chile}
\altaffiltext{10}{Astrophysikalisches Institut Potsdam, An der Sternwarte 16, Potsdam 14482, Germany}
\altaffiltext{11}{Observatorio Astr\'{o}nomico Nacional SPM, Instituto de Astronom\'{i}a, Universidad Nacional Aut\'{o}noma de M\'{e}xico, Ensenada, BC, M\'{e}xico}
\altaffiltext{12}{Las Cumbres Observatory Global Telescope, 6740 Cortona Dr. Ste. 102, Goleta, CA - 93117}
\altaffiltext{13}{Sydney Institute for Astronomy, School of Physics, The University of Sydney, NSW 2006, Australia}
\altaffiltext{14}{Department of Astronomy, University of Cape Town, Rondebosch 7700, South Africa}
\altaffiltext{15}{South African Astronomical Observatory, Observatory 7935, Cape Town, South Africa}
\altaffiltext{16}{Southern African Large Telescope Foundation, Observatory 7935, Cape Town, South Africa}
\altaffiltext{17}{Department of Astronomy \& Astrophysics, Villanova University, Villanova, PA - 19085}

\email{anjum@astro.washington.edu}

\begin{abstract}
Nonradial pulsations in the primary white dwarfs of cataclysmic variables
can now potentially allow us to explore the stellar interior of these accretors using stellar seismology.
In this context, we conducted a multi-site campaign on the accreting pulsator \sdsosf\ (V386~Ser)
using seven observatories located around the world
in May 2007 over a duration of 11 days. We report the best fit periodicities here, which were also previously observed in 2004,
suggesting their underlying stability. Although we did not uncover
a sufficient number of independent pulsation modes for a unique seismological fit, our campaign revealed that 
the dominant pulsation
mode at 609\,s is an evenly spaced triplet. The even nature of the triplet is suggestive of rotational splitting,
implying an enigmatic rotation period of about 4.8 days. There are
two viable alternatives assuming the triplet is real:
either the period of 4.8 days is representative of the rotation period of the entire star with implications for the
angular momentum evolution of these systems, or it is perhaps an indication
of differential rotation with a fast rotating exterior and slow rotation deeper in the star. 
Investigating the possibility that a changing period could mimic a triplet suggests that this scenario
is improbable, but not impossible.

Using time-series spectra acquired in May 2009, we
determine the orbital period of \sdsosf\ to be 83.8\,$\pm$\,2.9\,min.
Three of the observed photometric frequencies from our May 2007 campaign appear to be linear combinations of
the 609\,s pulsation mode with the first harmonic of the orbital period at 41.5\,min. This is the first discovery
of a linear combination between nonradial pulsation and orbital motion for a variable white dwarf.
\end{abstract}

\keywords{stars: dwarf novae--stars: individual (\sdsosf)--(stars:) novae, cataclysmic variables--stars: oscillations (including pulsations)--(stars:) white dwarfs--stars: rotation}

\section{Introduction to accreting white dwarf pulsators}
Cataclysmic variables are interacting binary systems in which a late-type
star loses mass to an accreting white dwarf.
Photometric variations consistent with nonradial g-mode pulsations were first discovered
in the cataclysmic variable GW~Librae in 1998 \citep{WarneravanZyl98,vanZylet00,vanZylet04};
such pulsations had previously been observed only among the non-interacting white dwarf stars.
This discovery has opened a new venue of opportunity
to learn about the stellar parameters of accreting variable white dwarfs
using asteroseismic techniques \citep[e.g.][]{Townsleyet04}.
A unique model fit to the observed periods of the variable
white dwarf can reveal information about the stellar mass, core
composition, age, rotation rate, magnetic field strength, and
distance \citep[see the review papers][]{Winget98,WingetaKepler08,FontaineaBrassard08}.

There are now thirteen accreting pulsating white dwarfs known
\citep[see][]{vanZylet04,WoudtaWarner04,WarneraWoudt04,
Pattersonet05a,Pattersonet05b,Vanlandinghamet05,Araujo-Betancoret05,Gaensickeet06,Nilssonet06,Mukadamet07b,Pavlenko08,Pattersonet08}.
\citet{Szkodyet02a,Szkodyet07,Szkodyet10} are pioneering the effort to empirically establish the pulsational
instability strip for accretors and to test the theoretical framework laid
down by \citet{Arraset06}. The instability strip(s) for these pulsators has to be established
separately from the \zzc\ strip\footnote{Non-interacting hydrogen atmosphere (DA) white dwarfs are observed to
pulsate in a narrow instability strip located within the temperature range
10800--12300\,K for $\log~g\approx 8$ \citep{Bergeronet95,Bergeronet04,
KoesteraAllard00,KoesteraHolberg01,Mukadamet04,Gianninaset05}, and
are also known as the \zzc\ stars.} because accretion enriches their envelopes
with He and metals. This is distinct from the pure H envelope
of the non-interacting DA white dwarfs, where H ionization causes them to pulsate as \zzc\ stars.
\citet{Arraset06} find a H/HeI instability strip for
accreting model white dwarfs with a blue edge near 12000\,K for a 0.6\,\msun star, similar
to the \zzc\ instability strip. They also find an additional hotter instability strip
at $\approx$15000\,K due to HeII ionization for accreting model white dwarfs with a
high He abundance ($>$\,0.38).

The spectrum of an accreting pulsator includes prominent broad absorption lines
from the white dwarf as well as the central emission features from the accretion disk.
When the orbital period of a cataclysmic variable is $\sim$80-90\,min, it is
near the evolutionary orbital period minimum, where the rate of mass transfer is theoretically expected
to be the smallest $\sim10^{-11}$\,\msun/yr \citep{KolbaBaraffe99}.
Due to the low rates of mass transfer, the white dwarf is expected to be the source of
90\% of the optical light observed from these systems \citep{TownsleyaBildsten02}.
This makes it possible to detect white dwarf pulsations in these
cataclysmic variables.

Accreting pulsators have probably undergone a few billion
years of accretion and thousands of thermonuclear runaways. Studying these systems
will allow us to address the following questions: to what extent does accretion affect the
white dwarf mass, temperature, and composition and how efficiently is angular momentum
transferred to the core of the white dwarf. These systems are also
crucial in understanding the above effects of accretion on pulsations.

Asteroseismology can allow us to obtain meaningful mass
constraints for the pulsating primary white dwarfs of cataclysmic variables.
Previously any such constraints on the mass of the accreting white dwarf could only be established 
for eclipsing cataclysmic variables \citep[e.g.][]{Woodet89,Silberet94,Singet07,Littlefairet08}.
Constraining the population, mass distribution, and evolution of accreting white dwarfs
is also important for studying supernovae Type Ia systematics.
For example, \citet{Williamset09} show empirically that the maximum mass of white dwarf progenitors has to be at least 7.1\,\msun\, 
thus constraining the lower mass limit for supernovae progenitors.

\section{Motivation}
\citet{Szkodyet02b} deduced that \sdsosf\ (V386~Ser; hereafter \sdsos) is a cataclysmic variable
from early SDSS spectroscopic
observations \citep{Stoughtonet02,Abazajianet03}; SDSS has single-handedly led to a substantial
increase in the number of known cataclysmic variables.
Subsequently \citet{WoudtaWarner04} discovered photometric variations
in the light curve of \sdsos\ consistent with non-radial pulsations. They determined
two independent pulsation modes with periods near 607\,s and 345\,s, also finding their harmonics
and linear combinations in the data. Noting the amplitude modulation in the light curves,
\citet{WoudtaWarner04} concluded that the dominant mode was a multiplet, but were unable to resolve it.
They also determined the orbital period of 80.52\,min from the observed double-humped modulation in their
light curves.

We chose to target \sdsos\
for a multi-site asteroseismic campaign from eleven possibilities known then for the following
reasons. Previous
observations of the system had revealed two independent pulsation frequencies,
while several similar systems show no more than one pulsation frequency. Each independent
pulsation frequency serves as a constraint on the stellar structure; detecting a larger
number of frequencies is essential in obtaining a unique seismological fit.
\sdsos\ has an equatorial declination, making it easily accessible to observatories in both northern and southern hemispheres.
The objective of the multi-site participation is to keep the sun from rising on the target star \citep{Natheret90}; reducing the
gaps in the stellar data due to daytime increases the
contrast between the true frequency and its aliases, thus making multi-site observations more effective than single-site observations.

\section{Observations}
We acquired optical time-series photometry on the accreting white dwarf
pulsator \sdsos\ over a duration of 11 days in May 2007 using multiple
telescopes. Our multi-site campaign involved using the prime focus time-series photometer Argos \citep{NatheraMukadam04}
on the 2.1\,m Otto Struve telescope at
McDonald Observatory (MO), and the time-series photometer Agile
\citep{Mukadamet07a} on the 3.5\,m telescope at Apache Point Observatory (APO).
Both instruments are frame transfer CCD cameras devoid of mechanical shutters,
where the end of an exposure and the beginning of a new exposure is
triggered directly by the negative edges of GPS-synchronized pulses without any intervention from the data acquisition software.
There is also no dead time between consecutive exposures from CCD read times, making Argos and Agile ideal
instrumentation for the study of variable phenomena with millisecond timing accuracy.
Argos and Agile consist of back-illuminated CCDs with an enhanced back-thinning process
for higher blue quantum efficiency. Additionally, the E2V CCD\,47-20 in Agile
has an ultra-violet coating to enhance the
wavelength efficiency of the region 200--370\,nm to 35\%. As Apache Point Observatory is located
at an altitude of 2788\,m, we expect to detect at least some of
the blue photons in the range of 320--370\,nm.

We utilized the Calar Alto Faint Object Spectrograph (CAFOS) in imaging mode on
the 2.2\,m telescope at Calar Alto Observatory (CAO) during the campaign.
The duration between the FITS time stamp and the actual opening of the shutter
is expected to be a fraction of a second. Network Time Protocol (NTP) is used to discipline the data
acquisition computer, and we expect that the uncertainty in timing for a CAFOS image is of the order of 0.5\,s.
The SITe CCD in CAFOS has a quantum efficiency greater than 80\% in the wavelength range of 370--780\,nm.
Data was acquired on the robotic 2.0\,m Faulkes Telescope North (FTN) at Haleakala, Hawaii, using the instrument HawkCam1; this
telescope belongs to the Las Cumbres Observatory Global Telescope (LCOGT) network.
The data acquisition computer for HawkCam1 is synchronized to
a GPS time server. The delay between the UTSTART timestamp in the FITS images and the actual opening of the mechanical shutter
is expected to be of the order of a few milliseconds. This instrument contains an E2V CCD42-40, which is
thinned and back-illuminated for blue sensitivity.

\begin{figure}
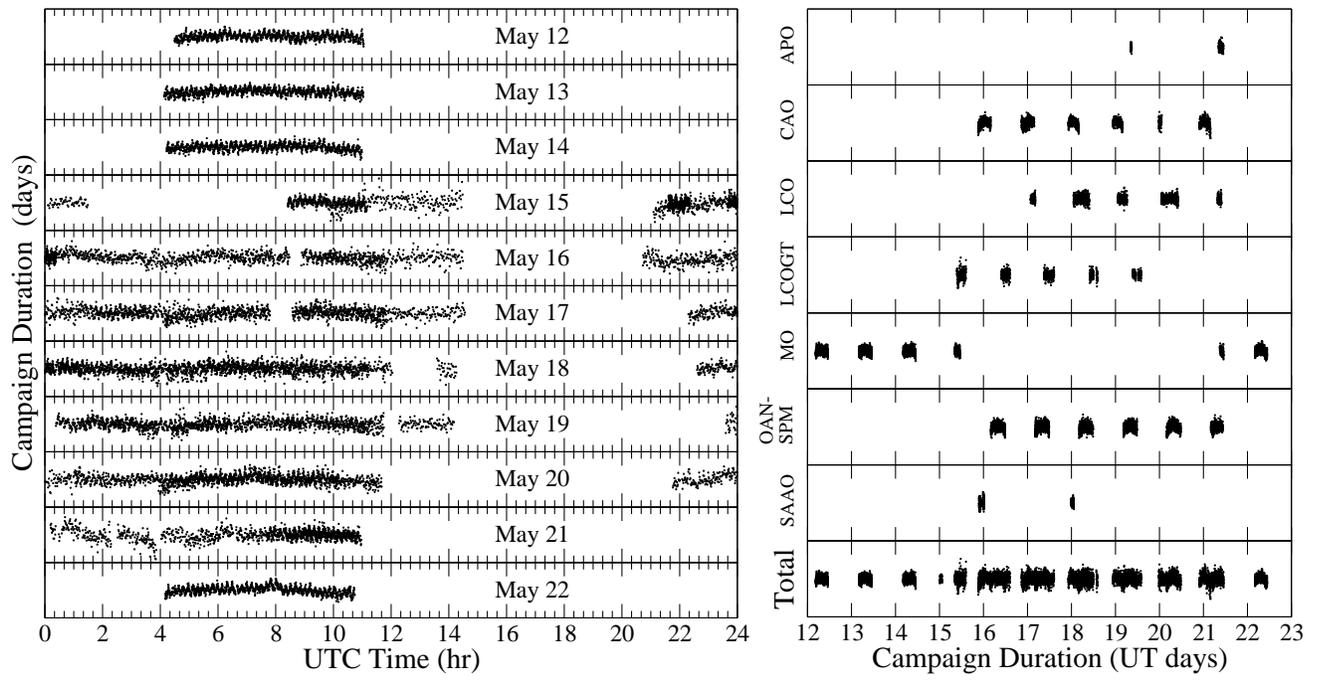

\includegraphics[height=3.5in,clip=true]{f1a.eps}
\includegraphics[height=3.5in,clip=true]{f1b.eps}
\caption{Total coverage obtained during the multi-site campaign.\label{xcov}}
\end{figure}

The camera SALTICAM \citep{O'Donoghueet03} was used on the effective 10\,m South African Large Telescope (SALT) at the South
African Astronomical Observatory (SAAO) to acquire data for our campaign. The frame transfer time for this large format
2K\,$\times$\,4K CCD is 0.1\,s; the uncertainty in timing should be significantly smaller than a tenth of a second.
The CCDs used in SALTICAM are thinned, back-illuminated,
and made from deep depletion silicon which provides less fringing and additional sensitivity in the near
infrared without photon loss in the blue and ultra-violet. SALTICAM
has an efficiency greater than 80\% in the wavelength range of 320--940\,nm. Including atmospheric extinction
and reflectivity losses at the different mirrors, SALTICAM has an efficiency greater than 60\% from 380 to 840\,nm.
We observed on the 1.5\,m telescope of the
Observatorio Astr\'{o}nomico Nacional in San Pedro Martir (OAN-SPM) using the instrument Ruca \citep{Zazuetaet00}.
Ruca comprises of the detector SITE1 SI003 CCD with a quantum efficiency greater than 55\% in the wavelength range 500--800\,nm.
On the 2.5\,m Ir\'{e}n\'{e}e du Pont telescope at
Las Campanas Observatory (LCO), we used the Direct CCD Camera ({\it CCD}), which includes a 2K\,$\times$\,2K Tek CCD with
a quantum efficiency greater than 70\% in the wavelength range of 400--700\,nm.
Table \ref{jour-obs} gives the journal of observations, and Figure \ref{xcov} shows the extent of our coverage during the campaign.

\begin{deluxetable}{llllllll}
\tabletypesize{\tiny}
\tablecolumns{8}
\tablewidth{0pc}
\tablecaption{Journal of Observations\label{jour-obs}}
\tablehead{
\colhead{Telescope} & \colhead{Instrument} & \colhead{Observers} & \colhead{Start Time} & \colhead{Ending Time} & \colhead{Number of} &\colhead{ExpTime} &\colhead{\hspace{-0.2in}Filter}\\
\colhead{}&\colhead{}&\colhead{}&\colhead{(UTC)}&\colhead{(UTC)}&\colhead{Images}&\colhead{(s)}&\colhead{}}
\startdata
APO 3.5\,m & Agile & ASM & 19 May 2007 08:17:03 & 08:55:33 & 77 &30 &BG40\\%AG0006
APO 3.5\,m & Agile & ASM & 21 May 2007 08:19:36.1 & 11:09:36.1 & 340 &30 &BG40\\%AG0007
CAO 2.2\,m & CAFOS & AA & 15 May 2007 00:05:49.0 & 01:29:47.5 & 56 & 75 & Roeser BV\\%CA_20070514
CAO 2.2\,m & CAFOS & AA & 15 May 2007 21:05:20.4 & 03:37:49.7 & 576 & 25 & none\\%CA_20070515
CAO 2.2\,m & CAFOS & AA & 16 May 2007 03:38:06.1 & 03:58:22.7 & 27 & 30 & none\\%CA_20070515
CAO 2.2\,m & CAFOS & AA & 16 May 2007 20:43:01.2 & 21:01:40.1 & 22 & 35 & none\\%CA_20070516
CAO 2.2\,m & CAFOS & AA & 16 May 2007 21:01:56.2 & 21:19:09.4 & 23 & 30 & none\\%CA_20070516
CAO 2.2\,m & CAFOS & AA & 16 May 2007 21:19:24.7 & 04:02:19.0 & 598 & 25 & none\\%CA_20070516
CAO 2.2\,m & CAFOS & AA & 17 May 2007 22:18:01.6 & 03:57:36.6 & 502 & 25 & none\\%CA_20070517
CAO 2.2\,m & CAFOS & AA & 18 May 2007 22:36:00.1 & 03:56:11.4 & 441 & 25 & none\\%CA_20070518
CAO 2.2\,m & CAFOS & AA & 19 May 2007 22:15:41.6 & 22:28:50.6 & 20 & 25 & none\\%CA_20070519
CAO 2.2\,m & CAFOS & AA & 19 May 2007 22:29:06.5 & 22:37:54.4 & 11 & 30 & none\\%CA_20070519
CAO 2.2\,m & CAFOS & AA & 19 May 2007 22:40:37.8 & 00:12:00.4 & 62 & 35 & none\\%CA_20070519
CAO 2.2\,m & CAFOS & AA & 20 May 2007 00:13:11.5 & 00:44:15.4 & 28 & 40 & none\\%CA_20070519
CAO 2.2\,m & CAFOS & AA & 20 May 2007 00:44:30.9 & 02:06:20.6 & 43 & 35 & none\\%CA_20070519
CAO 2.2\,m & CAFOS & AA & 20 May 2007 21:45:26.1 & 23:34:01.6 & 122 & 35 & none\\%CA_20070520
CAO 2.2\,m & CAFOS & AA & 20 May 2007 23:34:17.7 & 00:54:14.4 & 77 & 30 & none\\%CA_20070520
CAO 2.2\,m & CAFOS & AA & 21 May 2007 00:54:29.8 & 01:02:21.7 & 11 & 25 & none\\%CA_20070520
CAO 2.2\,m & CAFOS & AA & 21 May 2007 01:02:39.7 & 01:06:57.9 & 6 & 30 & none\\%CA_20070520
CAO 2.2\,m & CAFOS & AA & 21 May 2007 01:07:16.4 & 02:37:28.3 & 88& 35 & none\\%CA_20070520
CAO 2.2\,m & CAFOS & AA & 21 May 2007 02:37:44.6 & 03:29:59.7 & 63 & 30 & none\\%CA_20070520
CAO 2.2\,m & CAFOS & AA & 21 May 2007 03:30:15.0 & 03:40:49.2 & 16 & 25 & none\\%CA_20070520
CAO 2.2\,m & CAFOS & AA & 21 May 2007 03:41:12.2 & 03:51:36.1 & 14 & 30 & none\\%CA_20070520
LCO 2.5\,m & {\it CCD} & MRS, AS & 17 May 2007 01:39:26.4 & 03:14:38.1 & 101 & 40\tablenotemark{\alpha} & V\\%DP_20070517
LCO 2.5\,m & {\it CCD} & MRS, AS & 17 May 2007 03:15:51.3 & 04:01:04.7 & 61 & 30\tablenotemark{\alpha} & V\\%DP_20070517
LCO 2.5\,m & {\it CCD} & MRS, AS & 17 May 2007 04:01:45.4 & 04:31:19.0 & 46 & 25\tablenotemark{\alpha} & V\\%DP_20070517
LCO 2.5\,m & {\it CCD} & MRS, AS & 17 May 2007 04:34:20.4 & 05:42:11.8 & 86 & 35\tablenotemark{\alpha} & V\\%DP_20070517
LCO 2.5\,m & {\it CCD} & MRS, AS & 17 May 2007 05:46:22.7 & 06:10:28.1 & 25 & 45\tablenotemark{\alpha} & V\\%DP_20070517
LCO 2.5\,m & {\it CCD} & MRS, AS & 18 May 2007 01:05:50.7 & 02:20:54.2 & 101 & 30\tablenotemark{\alpha} & V\\%DP_20070518
LCO 2.5\,m & {\it CCD} & MRS, AS & 18 May 2007 02:21:43.3 & 05:32:14.0 & 293 & 25\tablenotemark{\alpha} & V\\%DP_20070518
LCO 2.5\,m & {\it CCD} & MRS, AS & 18 May 2007 05:32:50.8 & 09:05:45.5 & 362 & 20\tablenotemark{\alpha} & V\\%DP_20070518
LCO 2.5\,m & {\it CCD} & MRS, AS & 18 May 2007 09:06:06.7 & 09:25:37.6 & 31 & 25\tablenotemark{\alpha} & V\\%DP_20070518
LCO 2.5\,m & {\it CCD} & MRS, AS & 18 May 2007 09:26:11.3 & 09:51:10.2 & 35 & 30\tablenotemark{\alpha} & V\\%DP_20070518
LCO 2.5\,m & {\it CCD} & MRS, AS & 19 May 2007 01:16:31.5 & 08:32:53.0 & 650 & 25\tablenotemark{\alpha} & V\\%DP_20070519
LCO 2.5\,m & {\it CCD} & MRS, AS & 20 May 2007 01:02:52.3 & 10:11:13.3 & 843 & 25\tablenotemark{\alpha} & V\\%DP_20070520
LCO 2.5\,m & {\it CCD} & MRS, AS & 21 May 2007 01:59:23.6 & 02:42:51.1 & 62& 28\tablenotemark{\alpha} & V\\%DP_20070521
LCO 2.5\,m & {\it CCD} & MRS, AS & 21 May 2007 07:30:19.0 & 09:48:57.5 & 190 & 25\tablenotemark{\alpha} & V\\%DP_20070521
LCOGT 2.0\,m & HawkCam1 & Robotic & 15 May 2007 09:28:33.5 & 14:30:07.2 & 261 & 60 & V\\%FTN_20070514
LCOGT 2.0\,m & HawkCam1 & Robotic & 16 May 2007 09:42:41.3 & 14:29:40.8 & 244 & 60 & V\\%FTN_20070515
LCOGT 2.0\,m & HawkCam1 & Robotic & 17 May 2007 08:49:59.3 & 14:33:50.9 & 291 & 60 & V\\%FTN_20070516
LCOGT 2.0\,m & HawkCam1 & Robotic & 18 May 2007 09:53:09.2 & 12:03:08.0 & 113 & 60 & V\\%FTN_20070517
LCOGT 2.0\,m & HawkCam1 & Robotic & 18 May 2007 13:32:21.9 & 14:15:57.6 &38 & 60 & V\\%FTN_20070517
LCOGT 2.0\,m & HawkCam1 & Robotic & 19 May 2007 09:17:22.6 & 14:10:54.9 & 195 & 60 & V\\%FTN_20070518
MO 2.1\,m & Argos & ASM, BF & 12 May 2007 04:29:30.0 & 11:17:10.0 & 1223 & 20 &BG40\\%A1519
MO 2.1\,m & Argos & ASM, BF & 13 May 2007 04:07:39.0 & 11:10:19.0 & 1268 & 20 &BG40\\%A1521
MO 2.1\,m & Argos & ASM, BF & 14 May 2007 04:12:44.0 & 11:19:04.0 & 1279 & 20 &BG40\\%A1523
MO 2.1\,m & Argos & ASM, BF & 15 May 2007 08:24:49.0 & 11:12:49.0 & 504 & 20 &BG40\\%A1524
MO 2.1\,m & Argos & ASM, BF & 21 May 2007 08:58:08.0 & 10:58:08.0 & 240 & 30 &BG40\\%A1525
MO 2.1\,m & Argos & ASM, BF & 22 May 2007 04:09:51.0 & 11:00:39.0 & 1218 &20 &BG40\\%A1526
OAN-SPM 1.5\,m & Ruca & GT, SVZ & 16 May 2007 03:59:53 & 06:15:32 & 200 & 35 & BG40\\%OAN-SPM_20070516  
OAN-SPM 1.5\,m & Ruca & GT, SVZ & 16 May 2007 06:17:15 & 11:48:51 & 581 & 25 & BG40\\%OAN-SPM_20070516  
OAN-SPM 1.5\,m & Ruca & GT, SVZ & 17 May 2007 04:09:31 & 11:49:48 & 950 & 20 & BG40\\%OAN-SPM_20070517
OAN-SPM 1.5\,m & Ruca & GT, SVZ & 18 May 2007 04:12:23 & 11:47:23 & 799 & 25 & BG40\\%OAN-SPM_20070518
OAN-SPM 1.5\,m & Ruca & GT, SVZ & 19 May 2007 04:20:45 & 11:45:23 & 850 & 25 & BG40\\%OAN-SPM_20070519
OAN-SPM 1.5\,m & Ruca & GT, SVZ & 20 May 2007 03:56:09 & 11:40:19 & 892 & 25 & BG40\\%OAN-SPM_20070520
OAN-SPM 1.5\,m & Ruca & GT, SVZ & 21 May 2007 04:01:27 & 04:47:46 & 50 & 50 & BG40\\%OAN-SPM_20070521
OAN-SPM 1.5\,m & Ruca & GT, SVZ & 21 May 2007 04:50:12 & 10:39:22 & 500& 35 & BG40\\%OAN-SPM_20070521
SAAO 10\,m & SALTICAM & RS & 15 May 2007 21:36:18.6 & 00:23:41.3 & 605 & 3 & B-S1\tablenotemark{\beta}\\%SALT_20070515
SAAO 10\,m & SALTICAM & RS & 17 May 2007 23:49:51.7 & 01:15:55.5 & 287 & 10 & B-S1\tablenotemark{\beta}\\%SALT_20070517 
\enddata
\tablenotetext{\alpha}{The true exposure times are uneven and in excess of the given table entries
by 0.03--0.05\,s.}
\tablenotetext{\beta}{The B-S1 filter is nearly the same as a Johnson B filter.}
\end{deluxetable}

\section{Data Reduction and Analysis}
We used a standard IRAF reduction to extract sky-subtracted light curves
from the CCD frames using weighted circular aperture photometry \citep{O'Donoghueet00}.
After extracting the light curves, we divided the light curve of the target star with a sum of one or more
comparison stars using brighter stars for the division whenever available as opposed to faint stars; this choice
leads to comparatively lower noise in the target star light curve after the process of division.
After this preliminary reduction, we brought the data to a fractional amplitude
scale ($\Delta\,I/I$) and converted the mid-exposure times of the CCD images to Barycentric
Coordinated Time \citep[TCB;][]{Standish98}. The first data point of the campaign was used
to define a reference zero time of 2454232.693409 TCB, which we applied to
all the reduced light curves.
We then computed a Discrete Fourier Transform (DFT) for all the individual observing runs 
up to the Nyquist frequency.

All stellar pulsators reveal wavelength-dependent amplitudes whenever their
atmospheres suffer from limb-darkening effects; this is also observed to be true
for white dwarf pulsators \citep{Robinsonet95}.
\citet{Copperwheatet09} have measured the broadband amplitudes
of the pulsation frequencies in \sdsos, confirming the trend of larger amplitudes at bluer wavelengths.
Including photons redward of 6500\,\AA, which are
less modulated by the pulsation process, reduces the measured
amplitude by as much as 20--40\% \citep{Kanaanet00,NatheraMukadam04}.

Based on the pulsation amplitude of the dominant mode,
we divided the data from different sites, acquired
using different instruments with distinct wavelength responses and different filters,
into three groups. 
The data from the MO 2.1\,m, APO 3.5\,m, and the SAAO 10\,m telescopes along with the Roeser BV
data acquired on the 15th of May using the CAO 2.2\,m telescope constitute Group 1; these data
reveal an amplitude of 29.29$\pm$0.51\,mma for the 609\,s mode.
The data obtained from the LCO 2.5\,m, LCOGT 2.0\,m, and OAN-SPM 1.5\,m telescopes, that form Group 2,
yield an amplitude of 25.52$\pm$0.59\,mma. The Calar Alto white light
data that forms Group 3 amounts to an amplitude of 22.3$\pm$1.1\,mma.
These values of amplitude are not consistent with each other within the uncertainties.
The pulsation amplitude obtained for
sites in Group 2 is approximately 15\% lower compared to sites in Group 1, while
Group 3 shows a pulsation amplitude 31\% lower than Group 1. 
Hence to even out the 15--31\% amplitude differences between the various
data segments, we scaled the Group 2 data by a factor of 1.15 and Group 3 data by a factor of 1.31.
Without scaling the data, we would
effectively be fitting a constant amplitude sinusoid to a multi-site light curve with different embedded amplitudes.
After this scaling, the entire
11-day long light curve could be subjected to least squares analysis.

Blue bandpass filters are ideal to study these hot white dwarf variables.
Typically we are forced to acquire white light data whenever the photon count rate proves to be inadequate. Scaling these noisy
light curves to match the pulsation amplitude of the other higher S/N blue observations also implies
scaling the noise with the same factor. The drawback of scaling the data is an overall increase in noise; we do
find that the 3$\sigma$ level increases from 1.49\,mma to 1.71\,mma as a result of scaling. By comparison,
the highest amplitude
peak in the 609\,s triplet increases from 26.10\,mma to 28.78\,mma due to scaling.

We do not have to worry about scaling the other observed frequencies similarly, except the 41.5\,min period,
as they are all low amplitude modes.
Their amplitudes in the various data segments are already consistent with each other within the uncertainties.
We also checked explicitly that the chosen scaling does not alter the results of
the least squares analysis for any of the other frequencies, and no artificial frequencies
or frequency splittings arise due to this scaling.
Note that the amplitudes obtained for
the 41.5\,min period for sites in Group 1 and Group 2 are 5.22$\pm$0.50\,mma and 3.92$\pm$0.58\,mma respectively.
Sites in Group 2 show an amplitude approximately 33\% lower compared to sites in Group 1.
The wavelength dependence of the 41.5\,min period differs from the 609\,s mode; this is consistent with the
findings of \citet{Copperwheatet09}.
The Calar Alto white light
data (Group 3) are too noisy to determine a reliable amplitude for the 41.5\,min period.
This implies that our chosen scaling of 1.15 improves
the amplitude differences for the 41.5\,min period as well, and evens it out to the point that the data
segments have an amplitude consistent with each other within uncertainties.

\begin{figure}
\includegraphics[scale=0.65,clip=true]{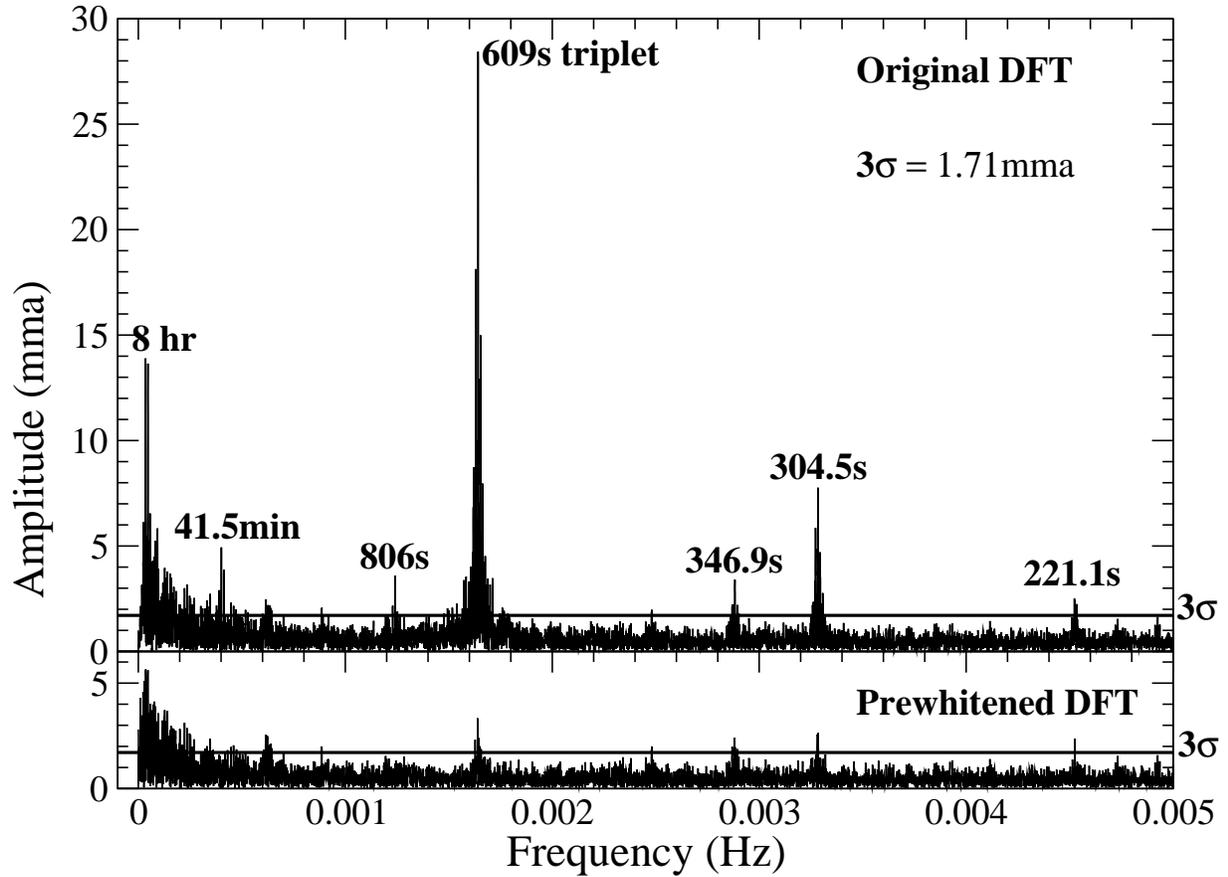}
\caption{Discrete Fourier Transform (DFT) of the multi-site data on \sdsos\ is shown in the top panel. The bottom panel
shows the prewhitened DFT obtained after subtracting all ten frequencies listed in Table 2.\label{DFT}}
\end{figure}

	All of the subsequent data analysis and the Discrete Fourier Transform (DFT) shown in Figure \ref{DFT}
is based on this scaled multi-site 11-day long light curve obtained
during the campaign. Although we computed the DFT right up to the Nyquist frequency of 0.025\,Hz,
there were no significant peaks above the 3\,$\sigma$ limit of 1.71\,mma\footnote{One milli modulation amplitude (mma) equals 0.1\% amplitude in intensity, corresponding
to a 0.2\% peak-to-peak change; one mma is equal to 1.085\,mmag.}
between 0.005 and 0.025\,Hz. Hence the interesting portion of the DFT is shown in Figure \ref{DFT}.
We determine the amount of white noise in the light curve empirically by using a shuffling technique \citep[see][]{Kepler93}.
All the best-fit frequencies were initially subtracted to obtain a prewhitened light curve.
Preserving the time column of this light curve, we shuffled the corresponding intensities. This exercise
destroys any coherent signal in the light curve while keeping the time sampling intact, leaving behind a shuffled light
curve of pure white noise. The average amplitude of a DFT computed for the shuffled light curve is close to the 1\,$\sigma$ limit of
the white noise. After shuffling the light curve ten times, we average the corresponding values for white noise to determine
a reliable 3\,$\sigma$ limit.

Table \ref{bfit-per} indicates the best fit periodicities along with the least squares uncertainties
we obtained using the program Period04 \citep{LenzaBreger05}.
We cross-checked the periods, amplitudes, and phases obtained with our own
linear and non-linear least squares code and find that the values obtained from both programs are almost
identical. The program Period04 has the added
advantage of computing Monte-Carlo uncertainties, which are shown as italicized numbers in Table \ref{bfit-per}.
Previous papers on \sdsos\ did not resolve the triplet \citep{WoudtaWarner04,Copperwheatet09}, and hence
their determinations of the dominant period differ slightly in value from ours (see the second to last paragraph in section 7).
However the pulsation spectrum in 2007 is essentially similar to that observed in 2004 and 2005.

\begin{deluxetable}{lllll}
\tablecolumns{5}
\tablewidth{0pc}
\tablecaption{Best Fit Periodicities: We show the observed periods and their
least squares uncertainties as well as italicized Monte Carlo uncertainties.\label{bfit-per}}
\tablehead{
\colhead{Identification} &
\colhead{Frequency ($\mu$Hz)} &
\colhead{Period (s)} &
\colhead{Amplitude (mma)} &
\colhead{Phase\tablenotemark{\alpha} (s)}}
\startdata
F$_{609\mathrm{a}}$ & 1641.796$\pm$ 0.028 &609.0893$\pm$ 0.010&28.78$\pm$ 0.50&596.0$\pm$ 2.2\\
& {\it $\pm$ 0.024} & {\it $\pm$ 0.0089} &{\it $\pm$ 0.63} & {\it $\pm$ 2.8} \\
\hline
F$_{609\mathrm{b}}$ & 1643.074 $\pm$ 0.057& 608.615$\pm$ 0.021 & 8.22$\pm$ 0.51& 255.0$\pm$ 6.1 \\
& {\it $\pm$ 0.057} &{\it $\pm$ 0.021} & {\it $\pm$ 0.49} & {\it $\pm$ 8.2} \\
\hline
F$_{609\mathrm{c}}$ & 1640.665$\pm$ 0.071&609.509 $\pm$ 0.026& 7.95 $\pm$ 0.66& 50.3 $\pm$ 6.9\\
& {\it $\pm$ 0.082} & {\it $\pm$ 0.031} & {\it $\pm$ 0.79} & {\it $\pm$ 9.1} \\
\hline
2\,F$_{609}$ & 3283.939  $\pm$ 0.040 &304.5124 $\pm$ 0.0037& 7.46$\pm$ 0.40& 132.0 $\pm$ 2.7\\
& {\it $\pm$ 0.053} & {\it $\pm$ 0.0049} &{\it $\pm$ 0.56} & {\it $\pm$ 3.3} \\
\hline
2\,F$_{609}$ & 3282.534 $\pm$ 0.073& 304.6427 $\pm$ 0.0068& 4.03 $\pm$ 0.40& 139.9  $\pm$ 5.0\\
& {\it $\pm$ 0.089} & {\it $\pm$ 0.0082} & {\it $\pm$ 0.52} & {\it $\pm$ 6.3} \\
\hline
F$_{806}$ & 1240.517 $\pm$ 0.071& 806.115 $\pm$ 0.046& 3.62 $\pm$ 0.40& 483  $\pm$ 14 \\
& {\it $\pm$ 10} & {\it $\pm$ 6.5} & {\it $\pm$ 1.4} & {\it $\pm$ 130} \\
\hline
F$_{221}$ & 4523.14 $\pm$ 0.10& 221.0854 $\pm$ 0.0049& 2.59 $\pm$ 0.40& 201.0 $\pm$ 5.4\\
& {\it $\pm$ 29} & {\it $\pm$ 1.4} & {\it $\pm$ 1.2} & {\it $\pm$ 78} \\
\hline
F$_{347}$ & 2882.582 $\pm$ 0.076& 346.9112 $\pm$ 0.0091& 3.41 $\pm$ 0.40& 91.6  $\pm$ 6.5\\
& {\it $\pm$ 23} & {\it $\pm$ 2.8} & {\it $\pm$ 1.5} & {\it $\pm$ 96} \\
\hline
F$_{41.5m}$ & 401.310  $\pm$ 0.053& 2491.84 $\pm$ 0.33 & 4.90 $\pm$ 0.40 & 826 $\pm$ 32\\
& {\it $\pm$ 0.075}  & {\it $\pm$ 0.46} & {\it $\pm$ 0.49} & {\it $\pm$ 45} \\
\hline
F$_{8h}$ & 34.766 $\pm$ 0.015 & 28764 $\pm$ 13 & 15.76  $\pm$ 0.43& 5860$\pm$ 110\\
& {\it $\pm$ 0.023}& {\it $\pm$ 19}& {\it $\pm$ 0.56}& {\it $\pm$ 130}\\
\enddata
\tablenotetext{\alpha}{We refer to the time of the first zero crossing of the sine curve as its phase, i.e. $\theta$=0
for a sine curve sin($\theta$) defined over the range from 0 to 2$\pi$. The phases are defined with respect to
a reference zero time of 2454232.693409 TCB.}
\end{deluxetable}

Note that we also concatenated all the light curves from various sites without any scaling and computed the
corresponding unscaled DFT, to find that it showed the same peaks as the DFT obtained after scaling the data. Additionally,
we computed DFTs for the unscaled light curves from the first and second groups as defined earlier. These DFTs
also revealed the same frequency set as shown in Figure \ref{DFT}. These tests
show that scaling the data did not introduce any artificial new frequencies in our data set.
In order to compute the best-fit periodicities for the unscaled light curve, we subjected it to Period04
similar to the procedure carried out for the scaled data. The best-fit values for the scaled and unscaled data
are very similar, lending credence to the process of scaling the multi-site data, especially in its ability to preserve the
original frequency set without introducing any new artifacts.

Although three of the seven sites boast millisecond timing accuracy, others like CAFOS involve a 0.5\,s timing
uncertainty per image. To test the effects of timing uncertainty on our results, we simulated a light curve
including all ten frequencies listed in Table \ref{bfit-per} along with their respective amplitudes,
adding in white noise with an amplitude of
2\,mma. Using the same time sampling as the real data, we added
in a Gaussian uncertainty of $\sigma$\,=\,2\,s to {\bf each} point to account for timing errors such as
software delays in time-shared data acquisition computers, jitter in the opening of mechanical shutters, etc.
This is worse than the worst case possible for badly synchronized clocks.
Our results show that we are still able to recover the closely spaced frequencies of the 609\,s triplet from the
simulated light curve; the output best-fit period values were at most 0.003\% different
from the values used to generate the simulated light curve. Amplitudes and phases were affected at
the 2\% level or less. This result is not surprising given that the timing
uncertainty is significantly smaller than the 609\,s period, and the S/N ratio improves with each passing cycle.

\section{Determining the orbital period}
\begin{figure}
\includegraphics[scale=0.8,clip=true]{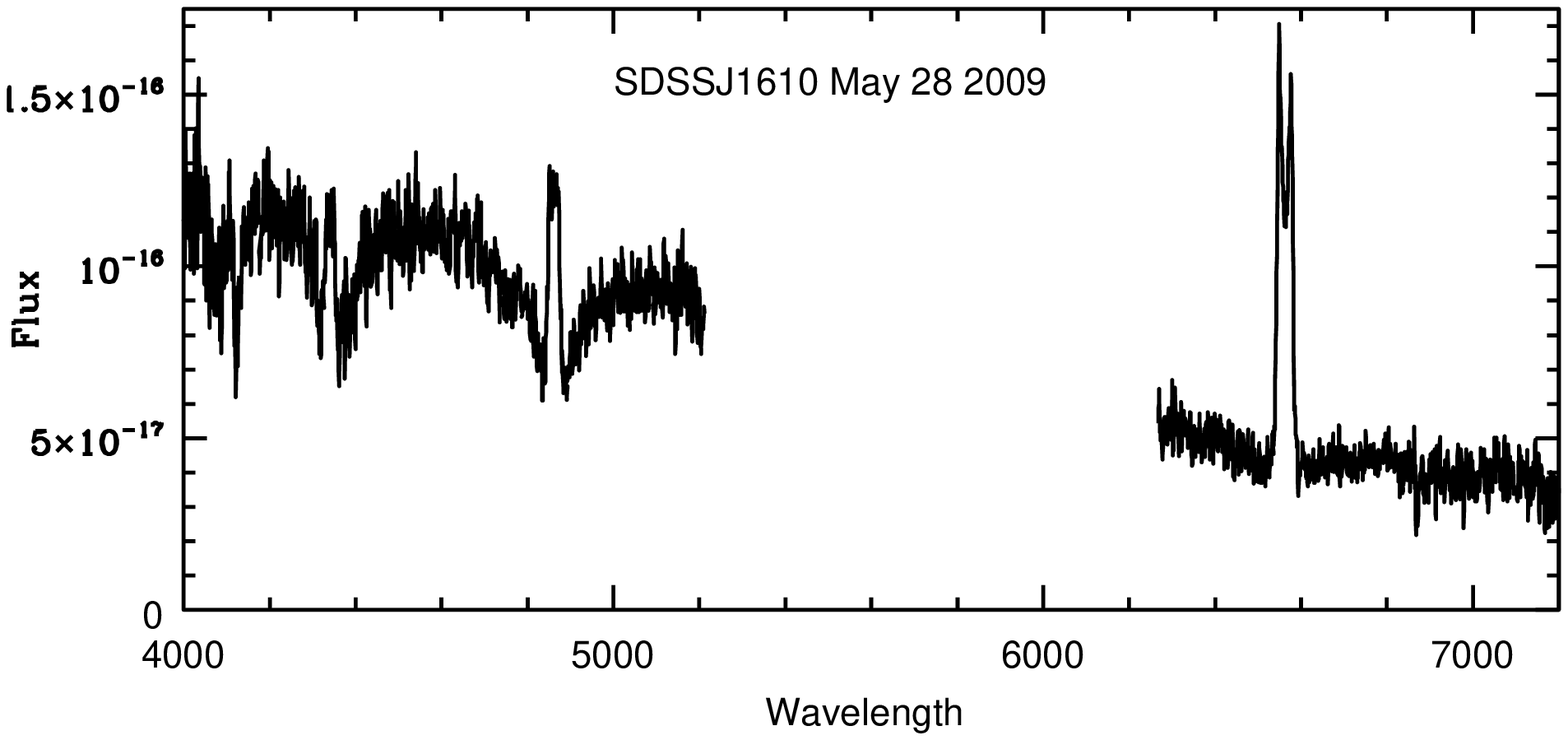}
\vspace{-3in}

\includegraphics[scale=0.65,angle=270,clip=true]{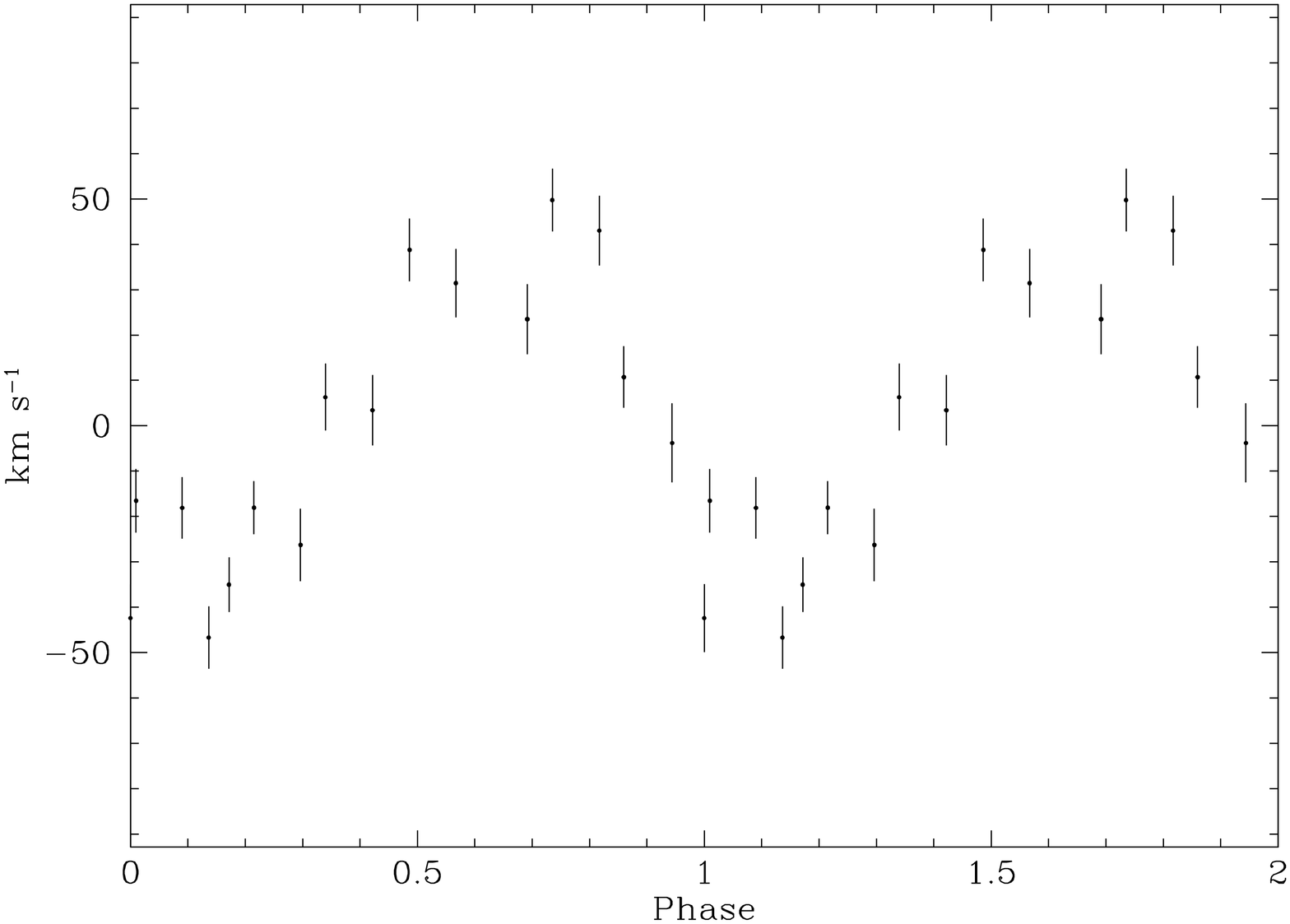}
\caption{The average spectrum of \sdsos\ obtained from 16 individual time-series spectra is shown in the top plot.
The radial velocity curve for the H$_{\alpha}$ line obtained using the double Gaussian method is shown
in the bottom panel.
\label{spec}}
\end{figure}

We acquired followup time-series spectroscopy on \sdsos\ on the 28th of May, 2009, using the
Dual Imaging Spectrograph (DIS) mounted on the APO 3.5\,m telescope with intermediate resolution
gratings (resolution about 2 \AA) and a 1.5 arcsec slit.
Our goal was not to get a precise determination of the
orbital period, but to measure the orbital frequency well enough to identify it or any of its harmonics
present in the campaign DFT. The spectroscopic data cover a period of
just about 3 hours, adding up to a total of sixteen 10-min exposures.
Individual exposures had a S/N of about 10 for the lines and 5--7 for the continuum.  
Standard IRAF routines were utilized for data reduction and subsequent flux and wavelength calibration.
We used a boxcar smoothing over 3 points prior to implementing
the centroid-finding routine in the IRAF
splot package to obtain the equivalent widths and fluxes for
the hydrogen Balmer lines.
Fitting a sine curve to the radial
velocity curves for each line, we determined the velocity of the center of mass $\gamma$,
the semi-amplitude K, the orbital period \porb, and the time of the red-to-blue crossing of the emission
lines T$_0$; these values are
shown below in Table \ref{spec-porb}. Note that the exposure times constitute a non-negligible part of the orbital period;
this causes some amplitude reduction. The amplitude values listed in Table \ref{spec-porb}
have been corrected for this effect.
The top panel of Figure \ref{spec} shows the complete combined spectrum.

\begin{deluxetable}{lllll}
\tablecolumns{5}
\tablewidth{0pc}
\tablecaption{Spectroscopic Orbital Period Measurement\label{spec-porb}}
\tablehead{
\colhead{Balmer} &
\colhead{Velocity of the Center} &
\colhead{Semi-amplitude} &
\colhead{T$_0$\tablenotemark{\alpha}} &
\colhead{Orbital Period}\\
\colhead{Line} &
\colhead{of Mass $\gamma$ (km/s)} &
\colhead{K (km/s)} &
\colhead{(min)} &
\colhead{\porb\ (min)}}
\startdata
H$_{\alpha}$ & -23\,$\pm$\,2 & 96\,$\pm$\,14\tablenotemark{\beta} & 410 & 83 \\
H$_{\beta}$  &  28\,$\pm$\,3 & 208\,$\pm$\,23\tablenotemark{\beta} & 412 & 86 \\
\enddata
\tablenotetext{\alpha}{The time of red-to-blue crossing of the emission lines T$_0$ is given in minutes
from 0\,hr UTC on 28 May 2009.}
\tablenotetext{\beta}{Since the exposure time was non-negligible compared to the orbital period,
the amplitude values have been corrected to account for the resulting reduction effect.}
\end{deluxetable}

Emission lines from the accretion disk and the hot spot, where the matter stream hits the
accretion disk, affect the spectral lines in cataclysmic variables. As a result, 
the velocity of the center of mass $\gamma$ can be different for different Balmer lines,
and may even have different signs (see Table \ref{spec-porb}).
However the determination of the orbital period from both lines is
consistent with each other within uncertainties.

We also applied the double Gaussian method outlined by
\citet{SchneideraYoung80} and developed by \citet{Shafter83} on the H$_{\alpha}$ line.
This method involved the convolution of two Gaussian functions of opposite signs and equal FWHM of 300\,km/s,
chosen to be the same order as the wavelength resolution. A peak separation of 1800\,km/s was used to reach further
into the wings without
compromising too much on the signal-to-noise ratio. This ensured that the measured radial velocities
traced the inner part of the accretion disc, where azimuthally symmetric structure traces the
orbital motion of the white dwarf.
This technique produced the radial velocity curve (bottom panel of
Figure \ref{spec}) that led to an orbital period determination of 83.8\,$\pm$\,2.9\,min.
This spectroscopic determination is consistent with the photometric measurement of 83.061\,$\pm$\,0.011\,min (obtained from 1/2\,F$_{41.5m}$)
within the uncertainties. This orbital period not only indicates that \sdsos\ is close to the period minimum, but also
includes \sdsos\ within the 80--86\,min period-minimum spike that is
largely made up of white dwarf dominated cataclysmic variables \citep{Gaensickeet09}.

\citet{WoudtaWarner04} observed a double-humped modulation in their
light curves, and determined values of 87.5\,$\pm$\,1.8\,min and 83.9\,$\pm$\,4.7\,min for the orbital period.
Using the first harmonic of the orbital period, they deduced additional
measurements of 81.9\,$\pm$\,1.2\,min and 81.2\,$\pm$\,0.4\,min; these photometric determinations have lower uncertainties 
compared to the direct measurements above inspite of their lower amplitude because they involve
twice the number of cycles compared to the orbital period itself.
Large gaps in the combined light curve prevented \citet{WoudtaWarner04} from identifying a unique orbital period, and they present
the alternatives of 80.52\,min and 85.08\,min. Computing a weighted average from their individual
measurements above, weighted
inversely as the uncertainties, we arrive at an orbital period determination of 83.7\,$\pm$\,1.4\,min.
This value is consistent with the orbital period measurement of 83.8\,$\pm$\,2.9\,min from the H$_{\alpha}$ line.

\begin{figure}
\includegraphics[scale=0.65,clip=true]{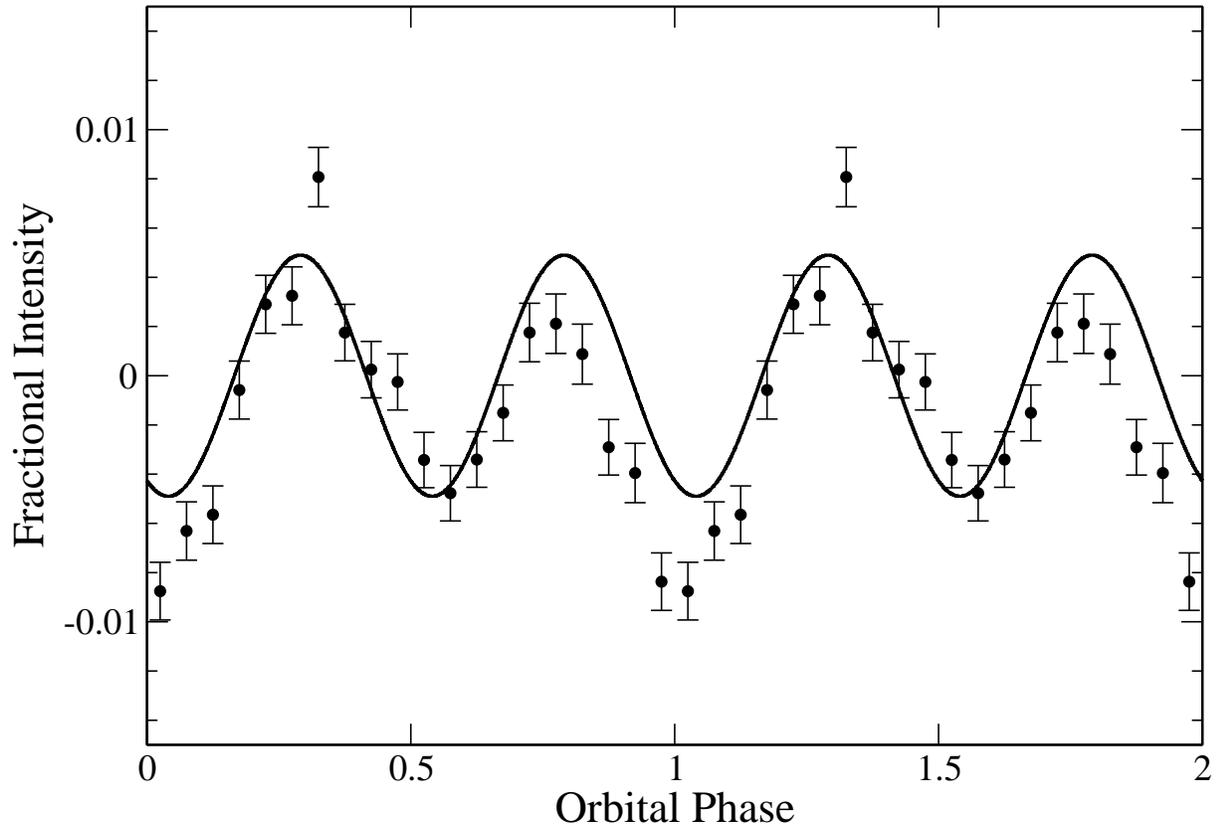}
\caption{Folding and binning the 11-day light curve on the orbital period after prewhitening all frequencies except the 41.5\,min periodicity
reveals a double-humped pulse shape (discrete points). We also show the sinusoidal fit listed in Table \ref{bfit-per} (continuous
line).\label{2hump}}
\end{figure}

We pre-whitened all the frequencies from our 11-day light curve except for the 41.5\,min periodicity, and folded it on the
orbital period. The folded light curve and its running average are shown
in Figure \ref{2hump} to indicate its apparent double-humped nature.
This pulse shape is different from the one shown by \citet{WoudtaWarner04}, but it is
consistent with the least squares analysis as expected. Table \ref{bfit-per} reveals that apart from the first harmonic,
no other period related to the orbital motion is observed. Hence we should expect to see two pulses with a period of about
41.5\,min at an amplitude of 4.9\,mma in the folded light curve. The pulse shape from Figure \ref{2hump} fits
a sine curve of period 41.24\,min with an amplitude of 4.3\,mma.

Several short period cataclysmic variables show similar double-humped light
curves \citep{Pattersonet98,RogozieckiaSchwarzenberg-Czerny01,Dillonet08},
and in some cases the orbital period has been independently determined using eclipses to confirm that the photometric and
spectroscopic periods are actually the same within uncertainties.
\citet{Robinsonet78} proposed that a double humped variation is observed when 
the bright spot, where the accretion stream hits the white dwarf, is visible even when it is behind the optically
thin disk. However \citet{Zharikovet08} and \citet{Avileset10} argue that the permanent double-humped light curve is
an attribute of bounced-backed systems; these systems consist of cataclysmic variables that have evolved beyond period minimum
with quiescent accretion disk radii at a 2:1 resonance.

The accreting pulsating white dwarf constitutes a clock in orbit; the changing light travel time during an
orbit is reflected in the pulse arrival times. The observed phase of the clock (O) compared to the phase
of a stationary clock at the same period (C) should show a modulation at the orbital period \citep{Wingetet03,Mullallyet08}.
We used the O-C method
(with a small modification) to attempt a detection of the orbiting motion of the 609\,s clock. However our uncertainties of
4--5\,s were too large to reveal the expected sub-second change in light travel time during the orbiting motion of the white dwarf.

\section{Identification of independent frequencies}
Each independent pulsation frequency is a constraint on the structure of the star. Harmonics and linear combinations
in our data merely arise as a result of non-linearities introduced by relatively thick convection
zones \citep{Brickhill92,Brassardet95,Wu01,Montgomery05}. Identifying the linearly independent frequencies
in a pulsation spectrum is the first step to undertake in seismology.

Of the several periods shown in Figure \ref{DFT}, we suspect that
the longest observed period of 28764\,s or 8\,hr is related to
airmass or extinction variations or possibly the gaps in the 11-day light curve.
Other cataclysmic variables have shown photometric or radial velocity periods that are much longer than the orbital
period and bear no relation to it. For example, \citet{WoudtaWarner02} discuss long photometric periods such as the
2\,hr modulation in GW\,Librae, a possible 3\,hr period in FS Aur, and a 4.6\,hr period in V2051 Oph.
\citet{Tovmassianet07} show that the precession of the magnetically accreting white dwarf can successfully explain
such long periods. However in this case, we do not believe that the 8\,hr period is stellar in origin.
The 2491.84\,s or the 41.5\,min period is clearly the first harmonic of the orbital period, measured spectroscopically
to be 83.8\,$\pm$\,2.9\,min. This identification is consistent with \citet{WoudtaWarner04}.

The 609\,s triplet is clearly an independent pulsation mode since it has the highest amplitude of all observed periods;
typically harmonics and linear combinations have lower amplitudes than the parent modes. We can also readily
identify the periods close to 304.5\,s as the first harmonic of the 609\,s triplet.
A triplet can be expected at 304.5\,s because the harmonic of a triplet
should also be a triplet. However we are unable to resolve it, due most likely to the relatively lower amplitudes
compared to the 609\,s triplet.

There is also a small chance that the 609\,s photometric period represents the rotation of the star,
and we are misinterpreting it as nonradial pulsation. \sdsos\ is neither strongly magnetic nor a source of x-rays,
and therefore we do not expect
to see a rotating hot spot as in the intermediate polars. It is difficult for a hot spot or belt to match the observed
pulse shape of the 609\,s period. If the 609\,s period was the rotation period of the star, it would also
be difficult to explain the evenly spaced triplet, but the triplet could
perhaps have been caused by reprocessing or reflection effects. The ratio of the UV to optical
pulsation amplitude of the 609\,s period is consistent with low order g-mode pulsations \citep{Szkodyet07}.
For all of these reasons collectively, nonradial pulsation is certainly the
favourable model to explain the 609\,s photometric triplet. However please note that we are currently
unable to absolutely exclude the possibility
that the 609\,s period is the rotation period of the star. For the rest of this paper, we will adopt the
more likely model of nonradial pulsation for the 609\,s triplet.

        Let F$_{41.5\mathrm{m}}$, F$_{806}$, F$_{609}$, F$_{347}$, and F$_{221}$ denote the frequencies
corresponding to the 41.5\,min, 806\,s, 609\,s, 347\,s, and 221\,s periods.
The linear combinations in our frequency set are: F$_{221}$\,=\,3\,F$_{609}$\,-\,F$_{41.5\mathrm{m}}$,
F$_{347}$\,=\,2\,F$_{609}$\,-\,F$_{41.5\mathrm{m}}$,
and F$_{806}$\,=\,F$_{609}$\,-\,F$_{41.5\mathrm{m}}$. Note that the
amplitudes of all the proposed combination modes are smaller than the suggested parent modes as expected.
Justifying the frequency F$_{221}$ involves invoking 3\,F$_{609}$, which has not been
observed directly. \citet{WoudtaWarner04} identify F$_{347}$ as a second independent pulsation mode; however given the relation
F$_{347}$\,=\,2\,F$_{609}$\,-\,F$_{41.5\mathrm{m}}$ within uncertainties, we arrive at the inevitable conclusion
that it is merely a linear combination mode.

This is probably the first instance for pulsating white dwarfs
where linear combination frequencies have apparently emerged as a result of
an interplay between a non-radial pulsation mode F$_{609}$ and a harmonic of the orbital frequency F$_{41.5\mathrm{m}}$.
The idea of a physical interaction between nonradial pulsation and tides caused by orbital motion
may be relevant here. We defer a thorough investigation to a future theoretical paper.

\section{Scrutinizing the pulsation triplet at 609\,s}
\begin{figure}
\includegraphics[scale=0.65,clip=true]{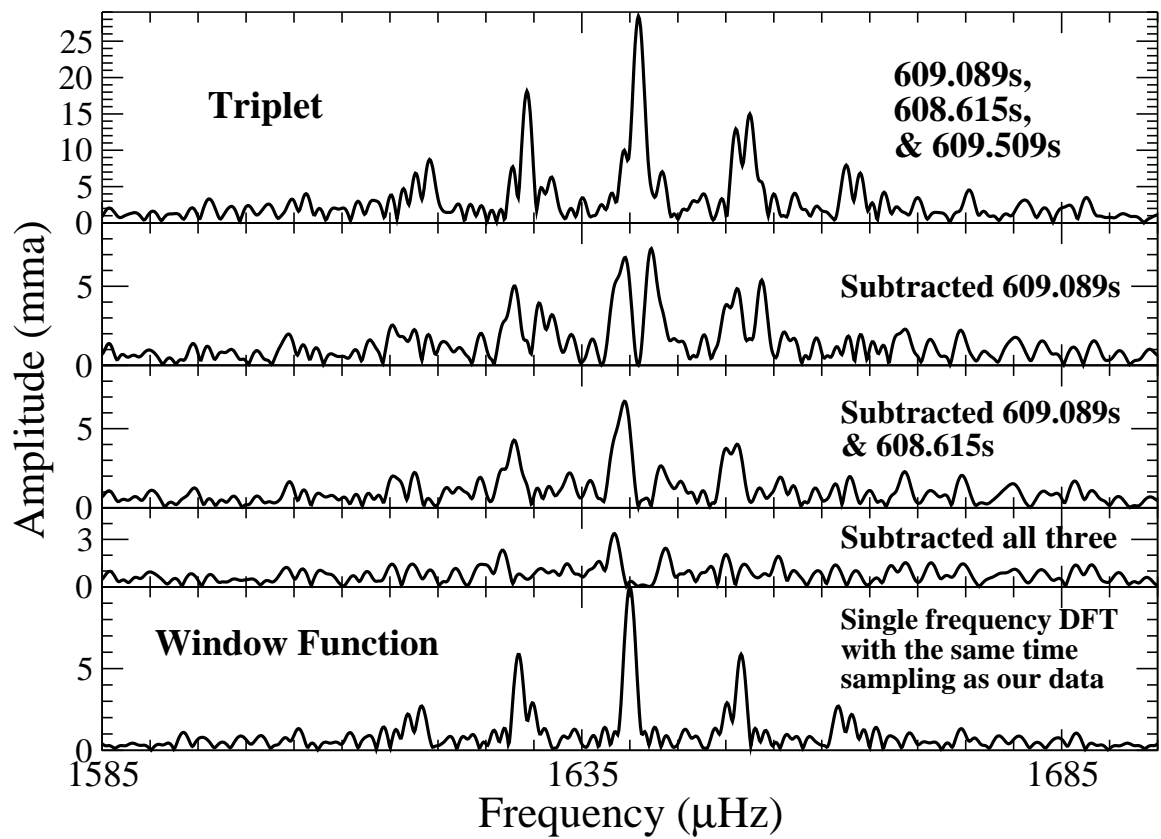}
\caption{The 609\,s dominant pulsation mode exhibited by \sdsos\ is a triplet. The top panel shows the original DFT, while subsequent
panels indicate the DFTs computed after subtracting the components of the multiplet one by one. The bottom panel shows the window function,
i.e. the DFT of a single frequency noiseless light curve with the same time sampling as the data.\label{trip}}
\end{figure}

A measure of the
gross ability of our data to resolve the components of the 609\,s mode is evident from
a comparison of the DFT and the window function, shown in the
top and bottom panels of Figure \ref{trip} respectively. The window function
is the DFT of a single frequency noiseless sinusoid sampled at exactly the same
times as the actual light curve. Comparing the DFT to the window function unambiguously reveals
the low frequency component and definite additional width, indicating the presence of more than one frequency.
Using least squares analysis in conjunction with the
technique of prewhitening, we are able to separate these resolved components.
Prewhitening is the standard procedure for isolating and characterizing closely spaced spectral
components; it involves subtracting out the best fit for the highest amplitude component of
the multiplet from the light curve and re-computing the DFT.
Carrying out this procedure for the 609.089\,s period gives clear residuals
with similar amplitude at similar spacing above and below in frequency, as
shown in the second panel of Figure \ref{trip}. Continuing to prewhiten with
the 608.615\,s (third panel) and then the 609.509\,s periods (fourth panel)
leads to marginal residual power near the 3$\sigma$ detection limit of
1.71\,mma.
This exercise demonstrates that the 609\,s mode is made up of three components in the form of a triplet
as listed in Table \ref{bfit-per}.

Measured pulsation amplitudes are typically lower than the intrinsic pulsation amplitudes
due to geometric
cancellation. This effect has three independent causes: disk averaging,
inclination angle, and limb darkening.
The inclination angle dictates the distribution of bright and dark
zones in our view for a given mode; this
introduces a large amount of scatter in observed pulsation amplitudes.
Eigenmodes with
different $m$ values exhibit different cancellation patterns (see \citealt{Dziembowski77,Pesnell85}).
Hence we did not expect that all members
of the 609\,s triplet should have had the same amplitude.

Given that the implications of the triplet spacing are most unexpected, we conducted simulations to verify whether
the triplet is indeed genuine or merely caused by a singular changing period. Although the pulsation
period of a non-interacting \zzc\ only drifts due to stellar cooling \citep{Kepleret00}, the period of an accreting
pulsator can drift on a faster timescale. During a dwarf nova outburst,
an accreting white dwarf is heated to temperatures well beyond the instability strip and it ceases to
pulsate \citep[e.g.]{Szkody08,Copperwheatet09,Szkodyet10}.
Once the white dwarf has cooled down close to its quiescent
temperature and pulsations have resumed in the star, \citet{Townsleyet04} explain how mode frequencies
can drift a little due to the continued cooling of the outer envelope.
They calculate that the longer period modes such as the 609\,s mode should drift at the rate of d$\nu$/dt\,$\sim$\,-$10^{-12}$\,Hz/s.
This expected model drift rate from the cooling of the outer envelope is much faster than the observed drift rate of
d$\nu$/dt\,$\sim$\,-$10^{-19}$\,Hz/s for \zzc\ pulsators from the cooling of the entire star \citep{Kepleret05,Mukadamet03,Mukadamet09}.
During our 11 day campaign, we should expect a drift in the pulsation period
of the order of 1\,$\mu$Hz, comparable to the observed triplet spacing. 
Hence it becomes necessary to check whether a drifting period is responsible for the observed triplet,
a fairly general way of introducing spurious peaks in the DFT.

	As a first step, we simulated constant-amplitude light curves with a single variable period near 609\,s; the period
was varied at different non-zero constant rates. The goal of the simulation was to check if it is possible to fit a triplet
to any of the generated light curves, while processing them in the same manner as the real light curve from the campaign.
For each drift rate dP/dt, we simulated a light curve
with exactly the same time sampling as the real data, adding in Gaussian noise
with an amplitude comparable to the observed white noise. We computed a DFT from the simulated
light curve, and then used the techniques
of least squares analysis and prewhitening to determine its frequency components, proceeding in exactly
the same way as the real data.

\begin{figure}
\includegraphics[scale=0.65,clip=true]{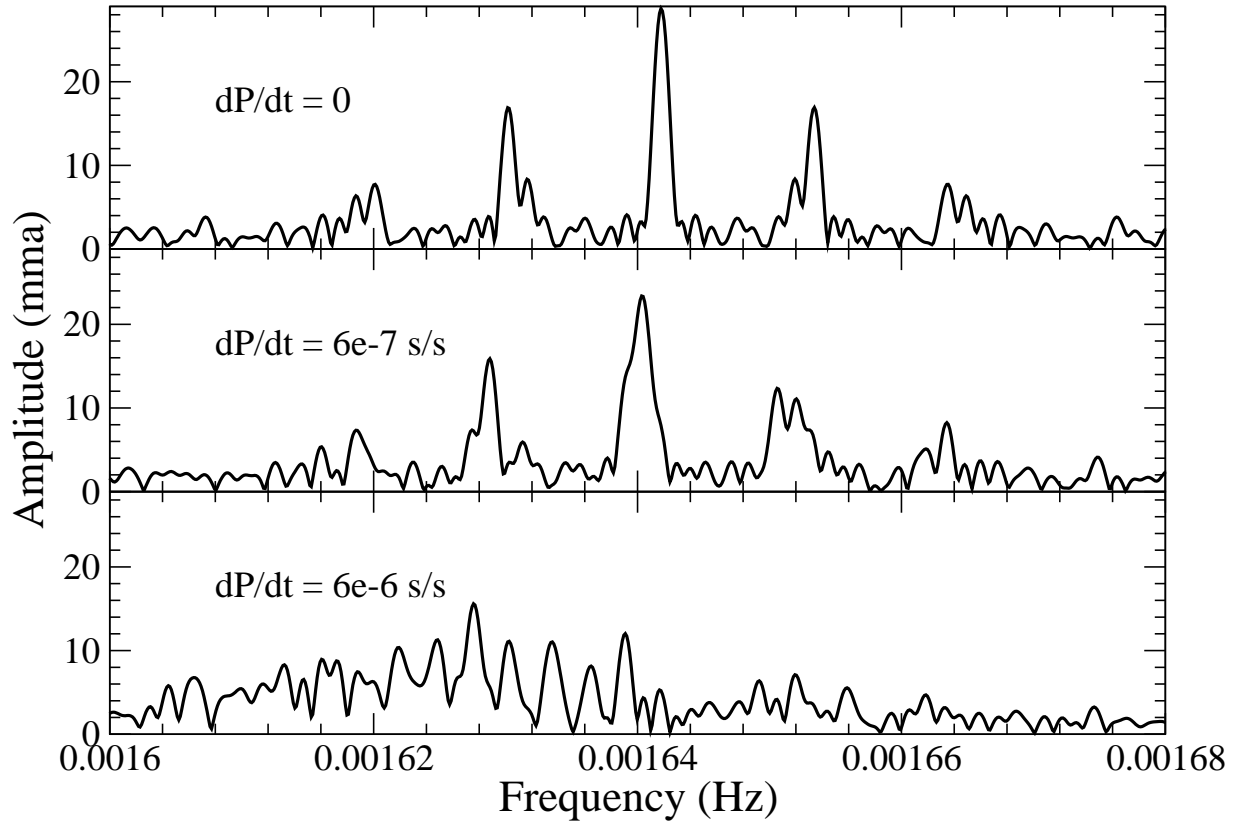}
\caption{We show DFTs of simulated mono-periodic constant-amplitude light curves with a time sampling identical
to the real data. The top panel reflects a constant period, while the middle and lower panels indicate
periods changing at the drift rates of dP/dt\,=\,6$\times10^{-7}$\,s/s and dP/dt\,=\,6$\times10^{-6}$\,s/s.
The bottom panel suggests that if the period varies too rapidly, then it is difficult to derive a coherent peak in the DFT.\label{Pdrift1}}
\end{figure}

Figure \ref{Pdrift1} shows the DFTs corresponding to the drift rates dP/dt\,=6\,$\times\,10^{-7}$\,s/s
and dP/dt\,=6\,$\times\,10^{-6}$\,s/s, also including the DFT of a single constant period at 609\,s for reference.
Figure \ref{Pdrift1} clearly demonstrates that the amplitude of the DFT based on the 11-day simulated
light curve is inversely proportional to the drift rate; faster the drift rate of the period,
smaller the amplitude of the variable period in the DFT. The drift rate of the period
has to be fast enough to cause a triplet spacing of 1.2\,$\mu$Hz in 11\,days,
and slow enough that the amplitude of the DFT does not fall significantly below the simulation amplitude.
We find that the drift rate of 6\,$\times\,10^{-7}$\,s/s comes close to reproducing the triplet spacing.
However we are unable to use non-linear least squares code to fit three closely-spaced
frequency components to the light curve simultaneously, indicating that they are not resolved.

\begin{figure}
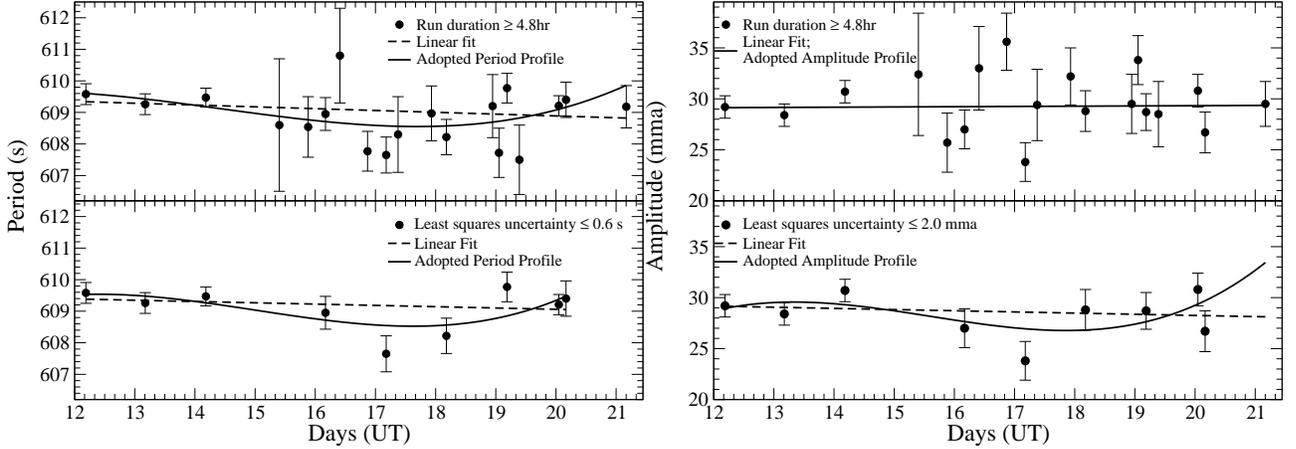

\includegraphics[scale=0.35,clip=true]{f7a.eps}\includegraphics[scale=0.35,clip=true]{f7b.eps}
\caption{The top panels show period and amplitude measurements of 19 runs longer than 4.8\,hr, made
assuming a single periodicity. The bottom panels concentrate on values with low uncertainties.
We show weighted first (dashed line) and third order polynomial (solid line) fits to the variations in period and amplitude.\label{dM}}
\end{figure}

	For the second step, we simulated mono-periodic signals whose
period and amplitude vary continuously as low order functions. 
In order to keep our simulations realistic, we determined possible low-order functions using the observed data.
To this end, we identified 19 runs of 4.8\,hr duration or longer,
acquired during the course of the campaign. Figure \ref{dM}
indicates our measurements of period and amplitude for these runs; the amplitudes
from filterless data are scaled by a factor of 1.31 and V filter data by a factor of 1.15.
Figure \ref{dM} shows a first and third order polynomial fit;
both sets of fits are weighted inversely as the squares of the uncertainties.
Points with uncertainties in period less than 0.6\,s and uncertainties in amplitude less than 2\,mma
are isolated in the lower panels of Figure \ref{dM}.
Note that apart from two points close to
the 17th and 18th of May, there is no strong evidence for any change in period.

For each low order function that fit the individual period measurements, we simulated light curves with all possible functions
that fit the amplitude measurements, to exhaust all possibilities of how the 609\,s periodicity could have changed.
In most cases, the amplitude of the DFT computed from the 11\,day simulated light curve fell
to values in the range of 15--20\,mma due to the varying period, thus making them less plausible.
It is crucial to note that the best fit to the amplitudes from the 19 long runs suggests that the amplitude stayed
constant at about 29\,mma, making a significant variation in amplitude unlikely.
Secondly, the amplitude of the combined DFT computed from the real data
is 28.8\,mma (see Table \ref{bfit-per}), at par with the individual amplitude measurements. This strongly suggests
that a substantial change in period during the 11-day campaign is also unlikely, as otherwise we would have observed
a lower amplitude in the combined DFT shown in Figure \ref{DFT} compared to the nightly amplitude measurements.

The low order functions we adopted above were meant to provide a realistic idea of
how the period and amplitude could change over the duration of the campaign. However none of our simulations with these functions
could produce a triplet that matched the observed triplet with its spacing and high amplitude.
For a fraction of the simulations that seemed plausible due to a reasonably high amplitude principal component,
we also used the pre-whitening and least squares analysis
code to determine the individual frequency components. For all of these cases, a simultaneous 3-frequency fit to the entire light curve
using the non-linear least squares program failed to converge. This is encouraging as the lack of
convergence apparently implies that the non-linear least squares program has the ability to distinguish between a single variable period
and a multiplet. Since we are unable to run a successful simulation where
a single variable period mimics a triplet while being constrained by the observed criteria,
we can only conclude that this possibility is not very likely. However we do recognize
that we can only carry out a finite number of simulations and scenarios, and therefore our simulations do not completely rule out
the possibility that a variable periodicity could mimic a triplet. We strongly recommend another campaign on this star
to see if a triplet with exactly identical components is observed again, and whether the future data stay in phase
with the observations presented here.

\begin{figure}
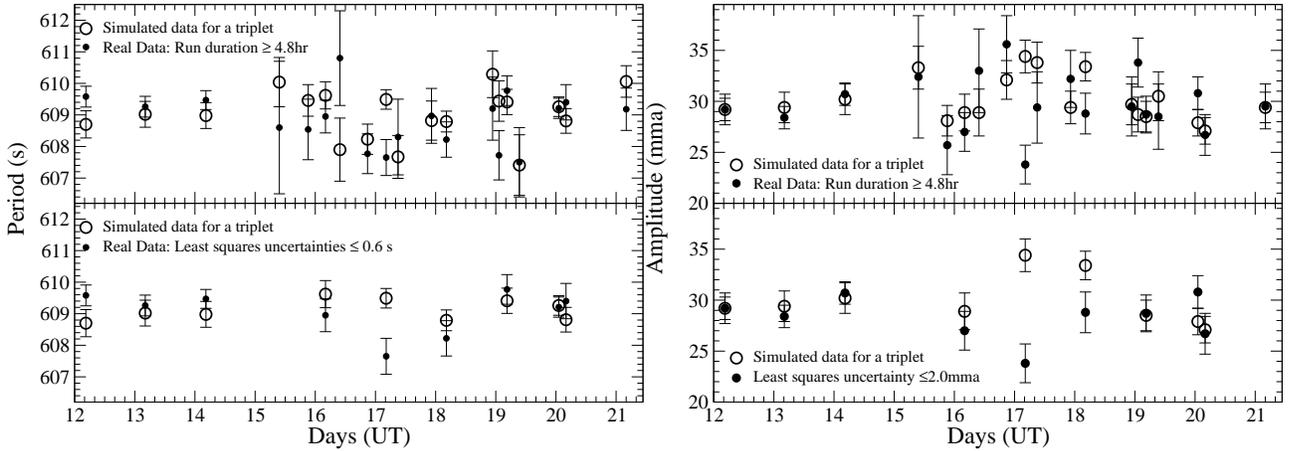

\includegraphics[scale=0.35,clip=true]{f8a.eps}\includegraphics[scale=0.35,clip=true]{f8b.eps}
\caption{The top panels show period and amplitude measurements of 19 runs longer than 4.8\,hr (small filled circles),
while the bottom panels concentrate on values with low uncertainties.
We also show the period and amplitude measurements for a simulated light curve (large hollow circles)
generated for a triplet with periods, amplitudes,
and phases from Table \ref{bfit-per} along with added Gaussian noise. This figure helps us understand the fluctuations
we can expect in the measurements of period and amplitude from runs much shorter than the duration required to resolve
the triplet, when fitting a single frequency to the data.\label{dM2}}
\end{figure}

As a last step, we also simulated a light curve for a triplet using the best-fit periods, amplitudes, and phases listed in
Table \ref{bfit-per}. We isolated 19 segments from the light curve corresponding to the 19
long runs discussed earlier. Then we determined the value of the period and amplitude of the dominant frequency for those 19 simulated runs;
we indicate these values in Figure \ref{dM2} and show the corresponding observed values for comparison. This exercise
helps us understand what fluctuations in period and amplitude we can expect to see for the triplet model when we
fit a single frequency to runs shorter than the duration needed to resolve the triplet. This also explains why
\citet{WoudtaWarner04} and \citet{Copperwheatet09} measure different values of the dominant period in \sdsos. Any significant
departure from the triplet model would pose as evidence in favour of a variable period model. Apart from a single point close
to the 17th of May, there is no significant deviation in period from the triplet model. The observed
amplitudes on 17th and 18th of May also do not match the expected amplitudes due to the triplet model. Despite these points,
we conclude that the triplet model matches the observed values well within the uncertainties, and is most probably correct.

The accreting pulsator GW~Librae also showed unresolved multiplet structure \citep{vanZylet04} with 1\,$\mu$Hz splitting
for the 370\,s and 650\,s modes. \citet{Araujo-Betancoret05} found similar unresolved multiplet structure
for the 336.7\,s mode of the accreting pulsator V455~And (HS\,2331+3905). Thus the multiplet spacing we observe
for \sdsos\ is neither unique nor unusual for accreting white dwarf pulsators; however this is the first data set
in which a multiplet has been cleanly resolved.

\section{Implications of the triplet spacing}
Each of the pulsation eigenmodes that can be excited in the star is typically
described by a set of indices: $k$, $\ell$, \& $m$.
The radial quantum number $k$ gives the number of nodes in the radial eigenfunction, and
$\ell$ and $m$ index the spherical harmonic $Y_{{\ell},m}$ function forming the
angular dependence of the mode.
Spherical harmonics (${{\mathrm{Y}}^{\ell }}_m$) are used to describe these eigenmodes due to the
spherical gravitational potential; this quantization
is similar to the quantum numbers used to describe the state of an electron
bound by the spherical electrostatic potential of the nucleus.
These non-radial motions penetrate below the surface with reducing amplitude;
the radial eigenfunction, which depends on $k$ and mode trapping, dictates the pulsation amplitude at different depths.
Montgomery \& Winget (1999) and
Montgomery, Metcalfe, \& Winget (2003) showed that pulsations probe up to the inner 99\% of a model white dwarf.

Multiplets in pulsation modes indicate a loss of spherical symmetry,
caused by either stellar rotation or a magnetic field \citep{Hansenet77, Joneset89}.
In the absence of a prominent magnetic field or rotation, normal
modes with different $m$ values will have the
same frequency, although they represent different fluid motion patterns.
According to \citet{Joneset89}, frequency splitting in the presence of a magnetic field is given by the following equation:

\begin{equation}
\sigma_{klm}(B_0) = \sigma_{kl}(B_0\,=\,0) + \sigma_{klm}'
\end{equation}

where the frequency shift $\sigma_{klm}'$ depends on $m^2$ for a dipole magnetic
field. Hence
a given mode splits into $\ell\,+\,1$ frequencies with uneven spacing.
A triplet caused by a magnetic field would require $\ell\,=\,2$, and the
individual components of the multiplet would have
$m$ values of 0, 1, \& 2. Such a triplet could not be evenly spaced because
the frequency shift of each component depends on $m^2$. Since the
observed 609\,s triplet is evenly spaced within uncertainties,
we conclude that it is not caused by a magnetic field.
The even spacing of the 609\,s triplet is consistent with rotational splitting.
In light of the remarkable implications
this may have for the rotational properties of the star, a more careful
discussion of the significance of this result is necessary.

Since the ratio of the frequency splitting to the central frequency for this triplet is small,
it should be safe to use perturbation analysis to linear order,
except possibly in the case of differential rotation.
\citet{Hansenet77} show that the frequency splitting due to rotation in linear
perturbation theory is given by:

\begin{equation}
\sigma = {\sigma}_0 - m ( 1 - C - C_1) {\Omega}_0
\end{equation}

where $\sigma$ denotes the frequency in the rotating star,
${\sigma}_0$ the frequency of the corresponding mode in a
non-rotating star (with the same $k$, $\ell$, $m$), and ${\Omega}_0$ denotes
the solid body angular rotation frequency.  The uniform rotation co-efficient
$C$ depends on the radial and azimuthal order ($k,\ell$), while $C_1$,
relevant for non-uniform or differential rotation, also depends on $|m|$.
Initially ignoring the term $C_1$ allows us to arrive at a basic constraint on the rotation of the star,
but we will revisit differential rotation below.
Uniform rotation splits a mode of azimuthal quantum number $\ell$ into
$2\ell+1$ evenly spaced modes. \citet{Brickhill75} showed that
in the limit of high radial order $k\gg 10$, the co-efficient $C$ can be simplified
as follows:

\begin{equation}
C = \frac{1}{\ell (\ell + 1)}
\end{equation}

Using the mode structure of GW Lib as a guide
\citep{Townsleyet04}, the 609\,s mode likely has $k\,\sim\,15$ and so qualifies
marginally for the above approximation.
This formalism makes it possible to obtain a gross estimate for the implied
spin period of the white dwarf.
The spacing of the components of
the 609\,s triplet are 1.13\,$\pm$\,0.11\,$\mu$Hz and
1.278\,$\pm$\,0.085\,$\mu$Hz respectively. This spacing is consistent with
being even within uncertainties, and we adopt the spacing of
1.2\,$\pm$\,0.14\,$\mu$Hz. Assuming that we are seeing the
$m\,=\,1,\,0,\,-1$ components of an $\ell=1$ mode, the implied spin period is
$4.8\pm0.6$\,days.

\section{Discussion}
If even approximately correct, the spin period of 4.8 days is remarkably long for an
object accreting matter in a binary with an orbital period of 83.8 minutes.
The implied rotational velocity for a period of 4.8\,days would be $\leq$1\,km/s as opposed to the typical rotational
velocities of 300--400\,km/s observed for non-magnetic accreting white dwarfs \citep{Szkodyet05}.
The measured surface velocity (vsin\,i) for the cataclysmic variables VW Hyi is ~600\,km/s \citep{Sionet95},
WZ Sge is ~200--400\,km/s \citep{Longet03}, and U Gem is ~50--100\,km/s \citep{Sionet94}.  
\citet{Kepleret95} find the rotation period of the \zzc\ star G\,226-29 to be 8.9\,hr, while
\citet{Mukadamet09} compute the rotation period of \zzc\ itself to be 1.5 days; all known
\zzc\ stars are non-magnetic. Whether they pulsate or not, non-interacting white dwarfs in
general are known to rotate slowly \citep{Koesteret98,Bergeret05}.

Alternatively, perhaps the splitting of the 609\,s mode
is only indicative of the rotation period of the region of the star that it samples well.
In other words, we examine the possibility of differential rotation with a rapidly rotating exterior
and a slowly rotating interior.
\citet{PiroaBildsten04} find that the accreted angular momentum is shared with the
accumulated envelope on short timescales.
Given the small splitting, any
differential rotation must be small or constrained to a very localized layer
of the star (e.g. the surface). This again provides strong constraints on
any diffusive mechanism for angular momentum transport. Evaluating
the constraints quantitatively requires defining both the mode
eigenfunctions and candidate rotation profiles, as well as how the rotational shear depends on 
latitude. Such an extensive theoretical analysis is beyond the scope of
this observational paper, but will be pursued in the near future by members of
our collaboration.

Such a long spin period of 4.8 days derived from the triplet spacing cannot be
discounted out of hand. While white dwarfs gaining matter are thought to spin up
\citep[e.g.][]{Kinget91,YoonaLanger04}, those undergoing classical novae are thought to eject
any accreted mass (see discussion in \citealt{TownsleyaBildsten04}) or
lose small quantities of mass based
on the abundance in their ejecta (e.g. \citealt{Gehrzet98}).
Therefore these objects have ample opportunity, depending on
how angular momentum is exchanged between the accreted envelope and core, to
either gain or shed angular momentum and allow the core to spin up or down
over the several Gyr accretion history of old cataclysmic variables
like \sdsos. \citet{LivioaPringle98} propose a model in which the primary white dwarf
loses accreted angular momentum during nova outbursts.

The demonstration by \citet{CharbonnelaTalon05} of how internal gravity waves
can extract angular momentum from the solar core during its evolution
might provide a model for how the core of an accreting white dwarf
could be spun down, with the angular momentum carried away with the envelopes over many nova
ejections.
The transport of angular momentum in stably stratified layers within stars
remains poorly understood. Diffusive prescriptions, even ones which depend
on magnetic effects like that of \citet{Spruit02} used by
\citet{YoonaLanger04}, are contradicted by the observed rotational profile of
the Sun \citep{Thompsonet03}. Despite having been spun down on the main
sequence, there is no observed gradient in the rotation profile of the solar core, an
essential aspect of angular momentum transport in any diffusive prescription.
Additionally such prescriptions predict rotation periods much
shorter than those observed for isolated non-magnetic white dwarfs
\citep{Suijset08}.

\citet{Katz75} showed that a radial magnetic field in a partially crystallized white dwarf
would be sheared by differential rotation, leading to an increase in the azimuthal component
proportional to the cumulative angle of differential rotation. He also indicated that when
the field strength reached about $10^5$ Gauss, the star would be locked as a rigid rotator.
\citet{WarneraWoudt02} also demonstrate the need for a magnetic field for
rigid body rotation in a white dwarf.
Perhaps in the context of \sdsos\, this implies that its magnetic field
is weak.

\section{Results}
We conducted a multi-site campaign on the accreting pulsator \sdsos\ for a duration of 11 days using
seven observatories around the globe in May 2007. The photometric periods in our light curve
were consistent with previous observations from 2004 and 2005 \citep{WoudtaWarner04,Copperwheatet09}, indicating
their long-term stability.
The most interesting result from our multi-site campaign is
the detection of a resolved evenly spaced pulsational triplet at 609\,s.
The even spacing of the triplet suggests that it is induced by rotation, and the rotational period of 4.8 days
derived from the spacing has strong implications for the transport of angular momentum and its long-term evolution.
We investigated the possibility that variability of period
and/or amplitude in a single frequency light curve could produce the observed triplet, and find that this
is not a likely possibility.
Either the period of 4.8 days is a measure of the uniform rotation period for the entire star or it
is suggestive of differential rotation in the star.
In either case, the prospects of constraining rotation in an accreting white dwarf with asteroseismic techniques is immensely exciting.
Conducting similar multi-site campaigns on other accreting pulsators could
help us form a picture of how angular momentum is exchanged in the interior
of the white dwarf, and its significance from the perspective of binary
evolution and stability.
The rotation of the underlying white dwarf is also important for Type Ia
Supernovae \citep[e.g.][]{Howellet01,Wanget03,Piersantiet03,YoonaLanger05}.

Our spectroscopic measurement of the orbital period is 83.8\,$\pm$\,2.9\,min is consistent within uncertainties
with the photometrically observed first harmonic of the orbital period at 41.5\,min.
Our second striking result is the detection of linear combination frequencies apparently caused by an interplay of
the dominant pulsation mode at 609\,s and
the first harmonic of the orbital period. Such a physical interaction
between nonradial pulsation and orbital motion has never been detected before for variable white dwarfs,
and is perhaps suggestive of tides. A thorough investigation is left to a future theoretical paper.

\acknowledgments
Drs. Szkody and Mukadam thank NSF for financially supporting this project through the grant AST-0607840. Dr. Bildsten
is grateful to NSF for the grant AST-0707633. Drs. Tovmassian and Zharikov acknowledge the CONACyT grant 45847/A for supporting the 
observations made at OAN-SPM. Professors Woudt and Warner acknowledge
research funding from the University of Cape Town and from the
National Research Foundation.
Based on observations obtained with the Apache Point Observatory 3.5-meter telescope,
which is owned and operated by the Astrophysical Research Consortium and
on observations collected at the Centro Astron\'{o}mico Hispano Alem\'{a}n (CAHA) at Calar Alto,
operated jointly by the Max-Planck Institut f\"{u}r Astronomie and the Instituto de Astrof\'{i}sica de
Andaluc\'{i}a (CSIC).
Some of the observations reported in this paper were obtained with the
Southern African Large Telescope (SALT).

\newpage

\onecolumn

\begin{thebibliography}{}

\bibitem[Abazajian et al.(2003)]{Abazajianet03} Abazajian, K.~et al.\
2003, \aj, 126, 2081

\bibitem[Araujo-Betancor et al.(2005)]{Araujo-Betancoret05}
Araujo-Betancor, S., et al.\ 2005, \aap, 430, 629

\bibitem[Arras et al.(2006)Arras, Townsley, \& Bildsten]{Arraset06} Arras, P., Townsley,
D.~M., \& Bildsten, L.\ 2006, \apjl, 643, L119

\bibitem[Aviles et al.(2010)]{Avileset10} Aviles, A., et al.\ 
2010, \apj, 711, 389 

\bibitem[Berger et 
al.(2005)]{Bergeret05} Berger, L., Koester, D., Napiwotzki, R., Reid, I.~N., \& Zuckerman, B.\ 2005, \aap, 444, 565 

\bibitem[Bergeron et al.(1995)]{Bergeronet95} Bergeron, P., Wesemael, F., Lamontagne, R., Fontaine, G., Saffer, R.~A., \& Allard, N.~F.\ 1995, \apj, 449, 258

\bibitem[Bergeron et al.(2004)]{Bergeronet04} Bergeron, P.,
Fontaine, G., Bill{\` e}res, M., Boudreault, S., \& Green, E.~M.\ 2004,
\apj, 600, 404

\bibitem[Brassard et al.(1995)Brassard, Fontaine, \& Wesemael]{Brassardet95} Brassard, P.,
Fontaine, G., \& Wesemael, F.\ 1995, \apjs, 96, 545

\bibitem[Brickhill(1975)]{Brickhill75} Brickhill, A.~J.\ 1975, 
\mnras, 170, 405 

\bibitem[Brickhill(1992)]{Brickhill92} Brickhill, A.~J.\ 1992,
\mnras, 259, 529

\bibitem[Charbonnel \& Talon(2005)]{CharbonnelaTalon05} Charbonnel, C. \&
Talon, S.\ 2005, Science, 309, 2189

\bibitem[Copperwheat et al.(2009)]{Copperwheatet09} Copperwheat, C.~M., 
et al.\ 2009, \mnras, 393, 157 

\bibitem[Dillon et al.(2008)]{Dillonet08} Dillon, M., et al.\ 
2008, \mnras, 386, 1568 

\bibitem[Dziembowski(1977)]{Dziembowski77} Dziembowski, W.\ 1977,
Acta Astronomica, 27, 203

\bibitem[Fontaine 
\& Brassard(2008)]{FontaineaBrassard08} Fontaine, G., \& Brassard, P.\ 2008, \pasp, 120, 1043 

\bibitem[G{\"a}nsicke et al.(2006)]{Gaensickeet06} G{\"a}nsicke,
B.~T., et al.\ 2006, \mnras, 365, 969

\bibitem[G{\"a}nsicke et al.(2009)]{Gaensickeet09} G{\"a}nsicke,
B.~T., et al.\ 2009, \mnras, in press (arxiv:0905.3476)

\bibitem[Gehrz et al.(1998)]{Gehrzet98} Gehrz, R.~D., Truran, J.~W.,
Williams, R.~E., \& Starrfield, S.\ 1998, PASP, 110, 3

\bibitem[Gianninas et al.(2005)]{Gianninaset05} Gianninas, A.,
Bergeron, P., \& Fontaine, G.\ 2005, \apj, 631, 1100

\bibitem[Hansen et al.(1977)]{Hansenet77} Hansen, C.~J., Cox, 
J.~P., \& van Horn, H.~M.\ 1977, \apj, 217, 151 

\bibitem[Howell et al.(2001)]{Howellet01} Howell, D.~A., 
H{\"o}flich, P., Wang, L., \& Wheeler, J.~C.\ 2001, \apj, 556, 302 

\bibitem[Jones et al.(1989)]{Joneset89} Jones, P.~W., Hansen,
C.~J., Pesnell, W.~D., \& Kawaler, S.~D.\ 1989, \apj, 336, 403

\bibitem[Kanaan et al.(2000)]{Kanaanet00} Kanaan, A., O'Donoghue,
D., Kleinman, S.~J., Krzesinski, J., Koester, D., \& Dreizler, S.\ 2000,
Baltic Astronomy, 9, 387

\bibitem[Katz(1975)]{Katz75} Katz, J.~I.\ 1975, \apj, 200, 298

\bibitem[Kepler(1993)]{Kepler93} Kepler, S.~O.\ 1993, Baltic 
Astronomy, 2, 515 

\bibitem[Kepler et al.(1995)]{Kepleret95} Kepler, S.~O., et al.\ 
1995, \apj, 447, 874 

\bibitem[Kepler et al.(2000)]{Kepleret00} Kepler, S. O.,
Mukadam, A., Winget, D. E., Nather, R. E., Metcalfe, T. S., Reed, M. D.,
Kawaler, S. D., \& Bradley, P. A. 2000, \apjl, 534, L185

\bibitem[Kepler et al.(2005)]{Kepleret05} Kepler, S.~O., et al.\
2005, \apj, 634, 1311

\bibitem[King et al.(1991)]{Kinget91} King, A.~R., Wynn, G.~A., 
\& Regev, O.\ 1991, \mnras, 251, 30P 

\bibitem[Koester et 
al.(1998)]{Koesteret98} Koester, D., Dreizler, S., Weidemann, V., \& Allard, N.~F.\ 1998, \aap, 338, 612 

\bibitem[Koester \& Allard(2000)]{KoesteraAllard00} Koester, D.~\&
Allard, N.~F.\ 2000, Baltic Astronomy, 9, 119

\bibitem[Koester \& Holberg(2001)]{KoesteraHolberg01} Koester, D., \&
Holberg, J.~B.\ 2001, ASP Conf.~Ser.~226: 12th European Workshop on White
Dwarfs, 226, 299

\bibitem[Kolb 
\& Baraffe(1999)]{KolbaBaraffe99} Kolb, U., \& Baraffe, I.\ 1999, \mnras, 309, 1034 

\bibitem[Lenz 
\& Breger(2005)]{LenzaBreger05} Lenz, P., \& Breger, M.\ 2005, Communications in Asteroseismology, 146, 53

\bibitem[Littlefair et al.(2008)]{Littlefairet08} Littlefair, S.~P., 
Dhillon, V.~S., Marsh, T.~R., G{\"a}nsicke, B.~T., Southworth, J., Baraffe, 
I., Watson, C.~A., \& Copperwheat, C.\ 2008, \mnras, 388, 1582 

\bibitem[Livio 
\& Pringle(1998)]{LivioaPringle98} Livio, M., \& Pringle, J.~E.\ 1998, \apj, 505, 339 

\bibitem[Long et al.(2003)]{Longet03} Long, K.~S., Froning, 
C.~S., G{\"a}nsicke, B., Knigge, C., Sion, E.~M., 
\& Szkody, P.\ 2003, \apj, 591, 1172 

\bibitem[Montgomery \& Winget(1999)]{MontgomeryaWinget99} Montgomery, M.\
H., \& Winget, D.\ E.\ 1999, \apj, 526, 976

\bibitem[Montgomery, Metcalfe, \& Winget(2003)]{Montgomeryet03}
Montgomery, M.~H., Metcalfe, T.~S., \& Winget, D.~E.\ 2003, \mnras, 344,
657

\bibitem[Montgomery(2005)]{Montgomery05} Montgomery, M.~H.\ 2005,
\apj, 633, 1142

\bibitem[Mukadam et al.(2003)]{Mukadamet03} Mukadam, A.~S., et al.\ 
2003, \apj, 594, 961 

\bibitem[Mukadam et al.(2004)]{Mukadamet04} Mukadam, A.~S., Winget,
D.~E., von Hippel, T., Montgomery, M.~H., Kepler, S.~O., \& Costa,
A.~F.~M.\ 2004, \apj, 612, 1052

\bibitem[Mukadam et al.(2007a)]{Mukadamet07a} Mukadam, A.~S., Owen, R., 
\& Mannery, E.~J.\ 2007a, Bulletin of the American Astronomical Society, 38, 159 

\bibitem[Mukadam et al.(2007b)]{Mukadamet07b} Mukadam, A.~S., 
G{\"a}nsicke, B.~T., Szkody, P., Aungwerojwit, A., Howell, S.~B., Fraser, 
O.~J., \& Silvestri, N.~M.\ 2007b, \apj, 667, 433 

\bibitem[Mukadam et al.(2009)]{Mukadamet09} Mukadam, A.~S.,
\etal\ 2009, Proceedings of the 16th European White Dwarf Workshop in Barcelona, in press

\bibitem[Mullally et al.(2008)]{Mullallyet08} Mullally, F., Winget, 
D.~E., Degennaro, S., Jeffery, E., Thompson, S.~E., Chandler, D., 
\& Kepler, S.~O.\ 2008, \apj, 676, 573 

\bibitem[Nather et al.(1990)]{Natheret90} Nather, R.~E., Winget,
D.~E., Clemens, J.~C., Hansen, C.~J., \& Hine, B.~P.\ 1990, \apj, 361, 309

\bibitem[Nather \& Mukadam(2004)]{NatheraMukadam04} Nather, R.~E.~\&
Mukadam, A.~S.\ 2004, \apj, 605, 846

\bibitem[Nilsson et al.(2006)]{Nilssonet06} Nilsson, R., Uthas, H.,
Ytre-Eide, M., Solheim, J.-E., \& Warner, B.\ 2006, \mnras, 370, L56

\bibitem[O'Donoghue et al.(2000)]{O'Donoghueet00} O'Donoghue, D.,
Kanaan, A., Kleinman, S.~J., Krzesinski, J., \& Pritchet, C.\ 2000, Baltic
Astronomy, 9, 375

\bibitem[O'Donoghue et al.(2003)]{O'Donoghueet03} O'Donoghue, D., et 
al.\ 2003, \procspie, 4841, 465 

\bibitem[Patterson et al.(1998)]{Pattersonet98} Patterson, J., 
Richman, H., Kemp, J., \& Mukai, K.\ 1998, \pasp, 110, 403 

\bibitem[Patterson et al.(2005a)]{Pattersonet05a} Patterson, J.,
Thorstensen, J.~R., \& Kemp, J.\ 2005a, \pasp, 117, 427

\bibitem[Patterson et al.(2005b)]{Pattersonet05b} Patterson, J.,
Thorstensen, J.~R., Armstrong, E., Henden, A.~A., \& Hynes, R.~I.\ 2005b,
\pasp, 117, 922

\bibitem[Patterson et al.(2008)]{Pattersonet08} Patterson, J., 
Thorstensen, J.~R., \& Knigge, C.\ 2008, \pasp, 120, 510 

\bibitem[Pavlenko(2008)]{Pavlenko08} Pavlenko, E.\ 2008, arXiv:0812.0791

\bibitem[Pesnell(1985)]{Pesnell85} Pesnell, W.~D.\ 1985, \apj,
292, 238

\bibitem[Piersanti et al.(2003)]{Piersantiet03} Piersanti, L., 
Gagliardi, S., Iben, I.~J., \& Tornamb{\'e}, A.\ 2003, \apj, 583, 885 

\bibitem[Piro 
\& Bildsten(2004)]{PiroaBildsten04} Piro, A.~L., \& Bildsten, L.\ 2004, \apj, 610, 977 

\bibitem[Robinson et al.(1978)]{Robinsonet78} Robinson, E.~L., 
Nather, R.~E., \& Patterson, J.\ 1978, \apj, 219, 168 

\bibitem[Robinson et al.(1995)]{Robinsonet95} Robinson, E.~L.~et
al.\ 1995, \apj, 438, 908

\bibitem[Rogoziecki 
\& Schwarzenberg-Czerny(2001)]{RogozieckiaSchwarzenberg-Czerny01} Rogoziecki, P., \& Schwarzenberg-Czerny, A.\ 2001, \mnras, 323, 850 

\bibitem[Schneider 
\& Young(1980)]{SchneideraYoung80} Schneider, D.~P., \& Young, P.\ 1980, \apj, 238, 946 

\bibitem[Shafter(1983)]{Shafter83} Shafter, A.~W.\ 1983, \apj, 
267, 222 

\bibitem[Silber et al.(1994)]{Silberet94} Silber, A.~D., 
Remillard, R.~A., Horne, K., \& Bradt, H.~V.\ 1994, \apj, 424, 955 

\bibitem[Sing et 
al.(2007)]{Singet07} Sing, D.~K., Green, E.~M., Howell, S.~B., Holberg, J.~B., Lopez-Morales, M., Shaw, J.~S., \& Schmidt, G.~D.\ 2007, \aap, 474, 951 

\bibitem[Sion et al.(1994)]{Sionet94} Sion, E.~M., Long, K.~S., 
Szkody, P., \& Huang, M.\ 1994, \apjl, 430, L53 

\bibitem[Sion et al.(1995)]{Sionet95} Sion, E.~M., Huang, M., 
Szkody, P., \& Cheng, F.-H.\ 1995, \apjl, 445, L31 

\bibitem[Spruit(2002)]{Spruit02} Spruit, H.~C.\ 2002, \aap, 381, 923

\bibitem[Standish (1998)]{Standish98} Standish, E.\ M.\ 1998,
\aap, 336, 381

\bibitem[Stoughton et al.(2002)]{Stoughtonet02} Stoughton, C.~et al.\
2002, \aj, 123, 485

\bibitem[Suijs et al.(2008)]{Suijset08} Suijs, M.~P.~L., Langer, N.,
Poelarends, A.-J., Yoon, S.-C., Heger, A., \& Herwig, F.\ 2008, \aap, 481, L87

\bibitem[Szkody et al.(2002a)]{Szkodyet02a} Szkody, P.,
G{\"a}nsicke, B.~T., Howell, S.~B., \& Sion, E.~M.\ 2002a, \apjl, 575, L79

\bibitem[Szkody et al.(2002b)]{Szkodyet02b} Szkody, P., et al.\ 
2002b, \aj, 123, 430 

\bibitem[Szkody et al.(2005)]{Szkodyet05} Szkody, P., Sion, E.~M., 
G{\"a}nsicke, B.~T.\ 2005, White dwarfs: cosmological and galactic probes, ASSL, 332, 205 

\bibitem[Szkody et al.(2007)]{Szkodyet07} Szkody, P., et al.\ 
2007, \apj, 658, 1188 

\bibitem[Szkody(2008)]{Szkody08} Szkody, P.\ 2008, HST Proposal, 
11639 

\bibitem[Szkody et al.(2010)]{Szkodyet10} Szkody, P., et al.\ 
2010, \apj, 710, 64 

\bibitem[Thompson et al.(2003)]{Thompsonet03} Thompson, M.~J.,
Christensen-Dalsgaard, J., Miesch, M.~S., \& Toomre, J.\ 2003, \araa, 41, 599

\bibitem[Tovmassian et al.(2007)]{Tovmassianet07} Tovmassian, G.~H., 
Zharikov, S.~V., \& Neustroev, V.~V.\ 2007, \apj, 655, 466 

\bibitem[Townsley 
\& Bildsten(2002)]{TownsleyaBildsten02} Townsley, D.~M., \& Bildsten, L.\ 2002, \apjl, 565, L35 

\bibitem[Townsley \& Bildsten(2004)]{TownsleyaBildsten04} Townsley, D.~M.~\&
Bildsten, L.\ 2004, \apj, 600, 390

\bibitem[Townsley et al.(2004)]{Townsleyet04} Townsley, D.~M.,
Arras, P., \& Bildsten, L.\ 2004, \apjl, 608, L105

\bibitem[Vanlandingham et al.(2005)]{Vanlandinghamet05} Vanlandingham,
K.~M., Schwarz, G.~J., \& Howell, S.~B.\ 2005, \pasp, 117, 928

\bibitem[van Zyl et al.(2000)]{vanZylet00} van Zyl, L., Warner,
B., O'Donoghue, D., Sullivan, D., Pritchard, J., \& Kemp, J.\ 2000, Baltic
Astronomy, 9, 231

\bibitem[van Zyl et al.(2004)]{vanZylet04} van Zyl, L., et al.\
2004, \mnras, 350, 307

\bibitem[Wang et al.(2003)]{Wanget03} Wang, L., et al.\ 2003, 
\apj, 591, 1110 

\bibitem[Warner 
\& van Zyl(1998)]{WarneravanZyl98} Warner, B., \& van Zyl, L.\ 1998, New Eyes to See Inside the Sun and Stars, 185, 321 

\bibitem[Warner
\& Woudt(2002)]{WarneraWoudt02} Warner, B., \& Woudt, P.~A.\ 2002, \mnras, 335, 84

\bibitem[Warner 
\& Woudt(2004)]{WarneraWoudt04} Warner, B., \& Woudt, P.~A.\ 2004, IAU Colloq.~193: Variable Stars in the Local Group, 310, 382 

\bibitem[Williams et al.(2009)]{Williamset09} Williams, K.~A., 
Bolte, M., \& Koester, D.\ 2009, \apj, 693, 355 

\bibitem[Winget(1998)]{Winget98} Winget, D. E. 1998, Journal of Physics:
Condensed Matter, 10, 11247

\bibitem[Winget et al.(2003)]{Wingetet03} Winget, D.~E., et al.\ 
2003, Scientific Frontiers in Research on Extrasolar Planets, 294, 59 

\bibitem[Winget 
\& Kepler(2008)]{WingetaKepler08} Winget, D.~E., \& Kepler, S.~O.\ 2008, \araa, 46, 157 

\bibitem[Wood et al.(1989)]{Woodet89} Wood, J.~H., Horne, K., 
Berriman, G., \& Wade, R.~A.\ 1989, \apj, 341, 974 

\bibitem[Woudt 
\& Warner(2002)]{WoudtaWarner02} Woudt, P.~A., \& Warner, B.\ 2002, \apss, 282, 433 

\bibitem[Woudt \& Warner(2004)]{WoudtaWarner04} Woudt, P.~A., \&
Warner, B.\ 2004, \mnras, 348, 599

\bibitem[Wu(2001)]{Wu01} Wu, Y.\ 2001, \mnras, 323, 248

\bibitem[Yoon 
\& Langer(2004)]{YoonaLanger04} Yoon, S.-C., \& Langer, N.\ 2004, \aap, 419, 623 

\bibitem[Yoon 
\& Langer(2005)]{YoonaLanger05} Yoon, S.-C., \& Langer, N.\ 2005, \aap, 435, 967 

\bibitem[Zazueta et al.(2000)]{Zazuetaet00} Zazueta, S., et al.\ 
2000, Revista Mexicana de Astronomia y Astrofisica, 36, 141 

\bibitem[Zharikov et
al.(2008)]{Zharikovet08} Zharikov, S.~V., et al.\ 2008, \aap, 486, 505

\end{thebibliography}
\end{document}